\documentclass[english]{article}
\usepackage[T1]{fontenc}
\usepackage[latin9]{inputenc}
\usepackage[a4paper]{geometry}
\usepackage{url}
\usepackage{amsmath}
\usepackage{amssymb}
\usepackage{graphicx}
\usepackage{setspace}
\usepackage{cite}
\usepackage{hyperref}
\begin{document}

\title{Anomalous diffusion and long-range memory in the scaled voter model}
\author{Rytis Kazakevi\v{c}ius\thanks{email: \protect\href{mailto:rytis.kazakevicius@tfai.vu.lt}{rytis.kazakevicius@tfai.vu.lt}}, Aleksejus Kononovicius\thanks{email: \protect\href{mailto:aleksejus.kononovicius@tfai.vu.lt}{aleksejus.kononovicius@tfai.vu.lt};
website: \protect\url{http://kononovicius.lt}}}
\date{Institute of Theoretical Physics and Astronomy, Vilnius University}
\maketitle

\begin{abstract}
We analyze the scaled voter model, which is a generalization
of the noisy voter model with time-dependent herding behavior.
We consider the case when the intensity of herding behavior grows 
as a power-law function of time. In this case, 
the scaled voter model reduces to the usual noisy voter model, 
but it is driven by the scaled Brownian motion. 
We derive analytical expressions for the time evolution of 
the first and second moments of the scaled voter model. 
In addition, we have derived an analytical approximation of 
the first passage time distribution. By numerical simulation, 
we confirm our analytical results as well as show that 
the model exhibits long-range memory indicators despite being a Markov model.
The proposed model has steady-state distribution consistent 
with the bounded fractional Brownian motion, 
thus we expect it to be a good substitute model for 
the bounded fractional Brownian motion.
\end{abstract}

\section{Introduction}

In recent years, methods of statistical physics have been increasingly
applied to describe complex social phenomena using tools common to physics, 
such as stochastic differential equations (SDEs), or recently developed ones,
such as agent-based models (ABMs) \cite{Pereira2017}.
This emerging area where physicists use
statistical physics techniques to solve financial and economic problems
is called econophysics. Analysis of the empirical data from various economic
and financial systems has shown that, despite the abundance of proposed
models, there is still a lack of models that accurately reproduce
and explain the emergence of empirically observable statistical properties
\cite{Cristelli2012Fermi}. It remains unclear what behavioral characteristics
of the individual system components can reproduce the empirical properties inherent
to such processes and fundamentally explain their origin \cite{Cont2007}.

One of the aforementioned problems is the nature of the observable
long-range memory. Numerous empirical long-range memory indicators
are well established and widely used: power-law power spectral density (PSD)
or power-law autocorrelation function and power-law scaling of the mean
squared displacement (MSD) over time \cite{Beran2013}. Also,
long-range memory can be identified by using the rescaled range method,
detrended fluctuation analysis, and other methods \cite{Kazakevicius2021Entropy}.
However, it is often difficult to determine from any single aforementioned
statistical property which process is responsible for the emergence
of long-range memory, and empirical methods often yield contradicting
results. 
For example, power-law PSD can be observed in a variety of stochastic
processes: nonlinear transformations of the Markov process \cite{Gontis2012ACS}, Brownian
motion subordinated to the L\'evy noise \cite{Kazakevicius2015physA}, or the fractional Brownian motion (fBm).
The aforementioned models also exhibit other indicators
of long-range memory such as power-law scaling of MSD, i.e.,
the anomalous diffusion. Therefore is not clear which model is more
appropriate and justifiable for describing the relevant empirical
time series \cite{Beran2013,Gontis2017PhysA}.

Various attempts to solve this problem have been made. In Refs. \cite{Gontis2017PhysA,Gontis2017Entropy}
it has been shown that the  "true" long-range
memory process, one with correlated time increments, such as fBm,
can be distinguished from other Markov processes by studying their
first passage time distributions (FPTDs). In the case of fBm,
both FPTD and PSD power-law exponents depend on the Hurst parameter
and for nonlinear Markov processes, the FPTD power-law exponent remains
constant regardless of the PSD power-law exponent value.
However, this method also has drawbacks. So far, this method has only
been applied to one-dimensional processes, i.e., it was assumed that
the statistical properties of the time series could be replicated
using a single variable SDE. It has been observed that a two-variable
SDE system can generate a time series with unique properties. For example,
a single nonlinear SDE can generate signals having power-law PSD only
if the stationary distribution of the signal itself is also a power-law
function. However, the system of two nonlinear SDEs can generate signals
with power-law PSDs, with arbitrary stationary distribution \cite{Ruseckas2016JStatMech}.
Therefore, it would be desirable to refine this method and apply it
to long-range memory identification in more complex systems, which
are described by a system of coupled SDEs derived from multi-state ABMs
\cite{Kononovicius2013EPL,Zha2020FI,Granha2022PNAS}.

Knowing of FPTD and other statistical properties lets us discern various
long-range memory processes. For example, fBm and L\'evy walks both exhibit
anomalous diffusion and power-law FPTD. However, 
fBm and  L\'evy walks  exhibit power-law FPTDs with 
different exponents \cite{Rangarajan2000}. So we chose
to test whether these aforementioned properties enable us to differentiate
noisy voter models from other long-range memory processes. In comparison
to previous works \cite{Kononovicius2013EPL,Kononovicius2014EPJB,Kazakevicius2021PRE}
here we have assumed that the intensity of herding behavior 
is not a model parameter (constant in time), but a function of time.
We chose herding dependence in the form of a power-law function because
such an introduction leads to very similar behavior compared to the scaled
Brownian motion (SBM) for the small times. SBM has similar statistical
properties as fBm except for PSD. fBm has power-law PSDs with an exponent
dependent on the Hurst parameter while PSD of SBM is
always inversely proportional to a frequency square as
in the case of the classical Brownian motion \cite{Sposini2019}.

The assumption that the intensity of herding behavior is time-dependent is
quite common in the literature
\cite{Haas2013NAJEF,Babalos2015AE,Kaizoji2015JEBO,Stavroyiannis2019RBF}.
Yet often it is assumed to be a stochastic process, while here we assume
that herding behavior follows a deterministic power-law function.  While our
choice appears to be great oversimplification, it still might be correct
close to the critical moments of high uncertainty.  In Ref.
\cite{Preis2010JStatPhys}, it was shown that trading volume exhibits
scale-free (power-law) behavior close to the trend-switching points.  Also,
in Ref. \cite{Barabasi2005Nature}, an issue is raised that many models in
sociophysics and econophysics are Poissonian, with inter-event times being
exponentially distributed, however, the empirical data indicates that
inter-event time distributions ought to be power-laws. To achieve
this in Poissonian models, one would need to have event rates be
time-dependent and power-law distributed.

This paper is organized as follows. In Sec.~\ref{sec:nvm},
we briefly introduce the noisy voter model
and SBM and their relevant statistical properties
such as time-dependence of MSD and FPTD.
In Sec.~\ref{sec:Voter-time-dependent-herding},
we show that the one-dimensional noisy voter model with a time-dependent
herding intensity for small times can be approximated by the
Cox-Ingersoll-Ross (CIR) process \cite{Cox1985} with time-dependent
coefficients \cite{Masoliver2016}. Additionally, general expressions
for first and second moments and variance have been obtained. In the
special case when the herding intensity is a power-law function of
time, the exact expressions for the moments have been derived.
In Sec.~\ref{sec:FPTD}, we show that the considered model is a nonlinear transformation
of SBM in an external potential and its FPTD has
the same power-law tail as SBM. In Sec.~\ref{sec:Applications},
we provide some remarks on how the proposed model relates to the other ABMs.

\section{Noisy voter model and scaled brownian motion}
\label{sec:nvm}

The voter model is one of the key models in sociophysics
\cite{Castellano2009RevModPhys,Cristelli2012Fermi}. It and its numerous
variations are still explored from various theoretical and empirical points of view \cite{Jedrzejewski2019CRP,Redner2019CRP}. Our
earlier research on the voter model \cite{Kazakevicius2021Entropy}
focused on the noisy voter model. We have shown that it can exhibit
long-range memory phenomenon \cite{Ruseckas2011,Kononovicius2012} as well as
be applied to explain spatial heterogeneity of electoral
\cite{Kononovicius2017Complexity} and census data
\cite{Kononovicius2019CompJStat}. Similar observations on the applicability
of the noisy voter model were also made by other groups
\cite{Alfarano2008Dyncon,Braha2017PlosOne,Fenner2017,Mori2019PRE}. Recently
we have analyzed diffusive regimes present in the nonlinear transformations
of the noisy voter model \cite{Kazakevicius2021PRE} as well as diffusive
properties of individual agent trajectories in the context of
parliamentary attendance \cite{Kononovicius2021CSF}.

The noisy voter model can be formulated as a birth-death process with the
following transition rates:
\begin{equation}
\pi^{+} =\left(N-X\right)\left(r_{1}+hX\right),\quad 
\pi^{-} =X\left(r_{2}+h\left[N-X\right]\right),\label{eq:nvm-rates}
\end{equation}
where in the equation above
$\pi^{+}$ stands for the birth rate (increment of the
system state $X$) and $\pi^{-}$ stands for the death rate (decrement of the
system state $X$). Note that system state variable $X$ is confined in the $\left[0,
N\right]$ interval. Therefore, one could see these transition rates not as
generation and recombination, but instead as $N$ particles (agents)
switching between two states (e.g., active and passive, Republican and
Democrat, etc.). The particles can switch the states independently with idiosyncratic
transition rate $r_{i}$, and they may change states due to interaction with
other particles which occurs with herding behavior intensity $h$.

As all the transitions in the model influence just one particle, we can use
one-step process formalism \cite{VanKampen2007NorthHolland} to derive an SDE
approximating the discrete process in the thermodynamic limit. For
$x=\frac{X}{N}$, the following SDE can be derived:
\begin{equation}
dx = h\left[\varepsilon_{1}\left(1-x\right)-\varepsilon_{2}x\right]dt+\sqrt{2hx\left(1-x\right)}dW_{t}\, .\label{eq:sdex}
\end{equation}
Here we have introduced relative independent transition rate
$\varepsilon_{i}=\frac{r_{i}}{h}$ by effectively coupling interaction rate
$h$ to the time scale. Also, in the SDE above, $W_{t}$ is the uncorrelated
standard Wiener process and Eq.~(\ref{eq:sdex}) should be interpreted in the  It\^o sense.

It can be trivially shown that the steady-state distribution of
Eq.~\eqref{eq:sdex} is the Beta distribution. The exact steady-state probability
density function (PDF) is:
\begin{equation}
P_{st}\left(x\right)=\frac{\Gamma\left(\varepsilon_{1}+\varepsilon_{2}\right)}{\Gamma\left(\varepsilon_{1}\right)\Gamma\left(\varepsilon_{2}\right)}x^{\varepsilon_{1}-1}\left(1-x\right)^{\varepsilon_{2}-1}.
\end{equation}

Beta distribution is observed in socio-economic data related
to popularity of political candidates or parties, also religions and
languages
\cite{Ausloos2007EPL,Braha2017PlosOne,Fenner2017,Mori2019PRE,Raducha2018PlosOne,Raducha2020SciRep,Ausloos2021PhysA}.
Hence, it is a popular model to study from a theoretical perspective, and to
compare against existing data.

Recently various non-Markovian modifications were introduced into the voter
model, and a non-Markovian voter model was considered as an alternative to
the original, Markovian, voter model. Refs. \cite{Artime2018PRE,Peralta2020PhysA}
have considered the implications of the state aging, this mechanism leads to a
frozen discord state, while the original voter model is known to reach
a consensus state. Refs. \cite{Raducha2018PlosOne,Raducha2020SciRep}
considered the evolution of the interaction topology alongside the evolution
in individual particle states. This extension was applied to model
competition between languages and language dialects. Further in this paper,
we will attempt to imitate non-Markovian behavior without introducing actual
non-Markovian mechanisms. We will do so by introducing SBM,
which is used to imitate certain features of fBm,
into the noisy voter model.

\subsection{Scaled Brownian motion and first passage time}
\label{sec:SBM}

In a later section, we will use SBM 
to describe the stochastic dynamics of the noisy voter model.
SBM can mimic some statistical properties of fBm such as power-law
FPTD and power-law scaling of MSD. Therefore,
here, we discuss SBM and its statistical properties in
a more detail.

SBM is well studied in the context of anomalous diffusion \cite{Szymanski2006,Wu2008}.
If MSD of observable $x$ has power-law
dependence on time, $\langle\left(\Delta x\right)^{2}\rangle=\langle x^{2}(t)\rangle-\langle x(t)\rangle^{2}\sim t^{\gamma}$,
then it is said that the process exhibits anomalous diffusion. Also if $\gamma\neq1$ one
can suspect that the process might exhibit long-range memory. If $\gamma<1$,
this phenomenon is subdiffusive. The occurrence of subdiffusion
has been experimentally observed, for example, in the behavior of
individual colloidal particles in random potential energy landscapes
\cite{Evers2013}, while the $1<\gamma\leq2$ case (known as superdiffusion) has been observed
in vibrated granular media \cite{Scalliet2015}. Recent research shows
that anomalous diffusion can be observed in socio-economic systems
\cite{Kazakevicius2021PRE}. For example, in Ref. \cite{Kononovicius2020JStatMech}
it was shown that anomalous diffusion can be observed by considering
individual agent trajectories in a modified voter model, thus providing
an explanation for the observations made in the parliamentary attendance
data \cite{Vieira2019PRE,Kononovicius2020JStatMech}.

SBM is one of the simplest Gaussian models that satisfies the anomalous
diffusion relation:
\begin{equation}
\langle x_{s}^{2}(t)\rangle\sim t^{\gamma_{s}},0<\gamma_{s}<2.
\end{equation}
Here $\langle x_{s}^{2}(t)\rangle$ is the MSD of SBM
and $\gamma_{s}$ is the anomalous diffusion exponent (SBM is a driftless
process, therefore its second moment coincides with MSD $\left\langle \left(\Delta x\right)^{2}\right\rangle =\langle x_{s}^{2}(t)\rangle$).
We have chosen to add subscript $s$ to the exponent to point out
that $\gamma_{s}$ is an anomalous diffusion exponent for SBM. In
a later section, we will see that the combination of SBM and the noisy
voter model can lead to a different anomalous diffusion exponents.

SDE describing SBM can be derived by rescaling the time of the Brownian
motion 
\begin{equation}
dx=\sqrt{2D}dW_{t}\, ,
\end{equation}
by using the following nonlinear time transformation $t\rightarrow
t_{s}=t^{\gamma_{s}}$,
leading to SDE in scaled time $t_{s}$ 
\begin{equation}
dx_{s}=\sqrt{2D}dW_{t_{s}}.\label{eq:W-process-in-scaled time}
\end{equation}

The equation above describes SBM in scaled time. Corresponding to SDE~\eqref{eq:W-process-in-scaled time} the Fokker-Planck equation is
\begin{equation}
\frac{\partial P(x_{s},t_{s})}{\partial t_{s}}=D\frac{\partial^{2}P(x_{s},t_{s})}{\partial x_{s}^{2}}.\label{eq:F-P-E-in-scaled-time}
\end{equation}
As we can see SBM, PDF $P\left(x_{s},t_{s}\right)$ in scaled
time $t_{s}$ satisfies the same Fokker-Planck equation as the Brownian
motion in real time $t$. Therefore, its PDF is identical in $t_{s}$
\begin{equation}
P(x_{s},t_{s})=\frac{1}{\sqrt{4D\pi t_{s}}}\exp\Bigg(-\frac{x_{s}^{2}}{4Dt_{s}}\Bigg),
\end{equation}
Returning from the scaled time to the real time, we can see that the PDF for SBM is
\begin{equation}
P(x_{s},t)=\frac{1}{\sqrt{4D\pi t^{\gamma_{s}}}}\exp\Bigg(-\frac{x_{s}^{2}}{4Dt^{\gamma_{s}}}\Bigg).\label{eq:Scaled-PDF}
\end{equation}

One can see that free SBM has the same PDF Eq.~(\ref{eq:Scaled-PDF})
as free fBm \cite{Vojta2020}. It has even been proposed that SBM is a possible substitute
for fBm in the large time limit \cite{Lim2002}.
Currently SBM is used to model anomalous diffusion in a wide range of systems
\cite{Szymanski2006,Wu2008}. However, it has been shown that due
to long-range correlations fBm PDF becomes different from SBM when
the boundary conditions or a potential are introduced \cite{Vojta2020,Guggenberger2021}.

The transition from scaled time $t_{s}=t^{\gamma_{s}}$ to real time
$t$ can be interpreted as a time derivative change in the Fokker-Planck equation 
\begin{equation}
\frac{\partial}{\partial
t_{s}}=\frac{1}{\gamma_{s}t^{\gamma_{s}-1}}\frac{\partial}{\partial t}.\label{eq:Time-deriv}
\end{equation}
By using Eqs.~(\ref{eq:Time-deriv}) and (\ref{eq:F-P-E-in-scaled-time})
we can obtain the Fokker-Planck equation describing SBM in real time
\begin{equation}
\frac{\partial P(x_{s},t)}{\partial t}=\gamma_{s}Dt^{\gamma_{s}-1}\frac{\partial^{2}P(x_{s},t)}{\partial x_{s}^{2}}.\label{eq:FPE-SBM-In-Real_Time}
\end{equation}
From Eq.~(\ref{eq:FPE-SBM-In-Real_Time}), it follows that SDE for
SBM in real time is
\begin{equation}
dx_{s}=t^{\frac{\gamma_{s}-1}{2}}\sqrt{2\gamma_{s}D}dW_{t}\, .\label{eq:SBM-in-real-time}
\end{equation}
Therefore, SBM can be interpreted as a Wiener process with a time-dependent diffusion
coefficient.

Now let us consider the first passage time denoted by
$T$, which is the time taken for the process to reach a threshold point $x=a$
for the first time, having started from the initial position $x_{0}$
at initial time $t=t_{0}=0$. Here $a$ is an absorbing boundary.
At absorbing boundary $x=a$ PDF $p(x,t|x_{0},0)$
must satisfy Dirichlet boundary condition $p(a,t|x_{0},0)=0$ for
all times.

The Fokker-Planck equation describing the Brownian motion
with a time-dependent diffusion coefficient, $\sigma^{2}(t)$, is
\begin{equation}
\frac{\partial p(x,t|x_{0},0)}{\partial t}=\frac{1}{2}\sigma^{2}(t)\frac{\partial^{2}p(x,t|x_{0},0)}{\partial x^{2}}.\label{eq:FPE-Wiener-Sigma(t)}
\end{equation}
For such type of Fokker-Planck equation first passage times, $T$,
distribution is well-known \cite{Molini2011,Bhatia2018}
\begin{equation}
f(T)=\frac{|x_{0}-a|}{2\sqrt{\pi}}\frac{e^{-\frac{(x_{0}-a)^{2}}{4S(t)}}}{S^{3/2}(T)}\frac{d}{dT}S(T),\label{eq:FPTD-wiener-Gen-Sigma-text}
\end{equation}
\begin{equation}
S(T)=\frac{1}{2}\int_{0}^{T}\sigma^{2}(t^{\prime})dt^{\prime}.\label{eq:Sigma-integral}
\end{equation}
Here $|x_{0}-a|$ is the absolute value of the difference between
the absorption point $a$ and the initial value of SBM. For the derivation
of Eq.~(\ref{eq:FPTD-wiener-Gen-Sigma-text}) see Appendix \ref{sec:FPTD-for-sigma-mu}.
By comparing Eqs.~(\ref{eq:FPE-SBM-In-Real_Time}) and
(\ref{eq:FPE-Wiener-Sigma(t)}),
we see that $\sigma^{2}(t)/2=\gamma_{s}Dt^{\gamma_{s}-1}$ therefore
\begin{equation}
S(T)=DT^{\gamma_{s}},\label{eq:S(T)}
\end{equation}
if $\gamma_{s}>0$. By inserting Eq.~(\ref{eq:S(T)}) into Eq.~(\ref{eq:FPTD-wiener-Gen-Sigma-text})
we obtain FPTD for SBM

\begin{equation}
f(T)=\frac{|x_{0}-a|\gamma_{s}}{2\sqrt{\pi D}}\frac{1}{T^{\gamma_{s}/2+1}}\exp\Bigg(-\frac{(x_{0}-a)^{2}}{4DT^{\gamma_{s}}}\Bigg).\label{eq:FPTD-SBM}
\end{equation}

For $\gamma_{s}=1$, Eq.~(\ref{eq:FPTD-SBM}) reduces to the well-known
FPTD for Brownian motion. As $t$ goes to $\infty$, FPTD $f_{s}(t)\sim1/t^{\beta}$
decays as a power-law function with exponent $\beta=\gamma_{s}/2+1$.
The aforementioned exponent is dependent on the anomalous diffusion parameter
$\gamma_{s}$. A special case of Eq.~(\ref{eq:FPTD-SBM}) for
absorption at origin $a=0$ ($x_{0}>a$) has been obtained by using
the method of mirrors \cite{Molini2011}. In addition, the FPTD for SBM
affected by the time-dependent force has been obtained in Ref. \cite{Molini2011}.
It is well-known that fBm also exhibits anomalous diffusion with MSD
$\langle x_{\text{fBm}}^{2}(t)\rangle\sim t^{2H}$ and has FPTD with a
power-law tail $f_{\text{fBm}}(T)\sim \frac{1}{T^{H+1}}$
\cite{Lim2002}. Here $H$ is the Hurst parameter. If we set $\gamma_{s}=2H$,
then we can see that SBM and fBm exhibit the same power-law scaling behavior 
in MSD and in FPTD. fBm has a power-law PSD dependent on the Hurst
exponent, however, SBM PSD is always proportional
to $1/f^{2}$ and does not depend on the anomalous diffusion exponent
\cite{Sposini2019}.

Other transformations of the stochastic processes can also lead to
anomalous diffusion. For example, the nonlinear transformation of
Brownian motion $y=x^{\eta}$ \cite{Cherstvy2013} or Bessel process
$y=(\eta-1)x_{\text{Bes}}^{1-\eta}$ \cite{Kazakevicius2016pre} or even
more complex transformations $y=(x_{V}/(1-x_{V}))^{1/\alpha}$ of
the noisy voter model \cite{Kazakevicius2021PRE} lead to the anomalous
diffusion. Here $x_{V}$ is the process defined by SDE~(\ref{eq:sdex})
and $x_{\text{Bes}}$ is defined by the Bessel process \cite{Gontis2012ACS}.
However, the aforementioned transformations do not change FPTD power-law
exponent $\beta$. The exponent remains equal to $3/2$ and independent
from the anomalous diffusion exponent \cite{Gontis2012ACS}. To obtain
the anomalous diffusion and power-law FPTD often non-Markovian processes are
used, such as fBm \cite{Ding1995fbm} or L\'evy walks \cite{Palyulin2019}.

Therefore, at least as far as we are aware, the SBM is the only Markovian
process exhibiting anomalous diffusion and power-law FPTD with an
exponent different from Brownian motion. Therefore we chose
SBM as a noise source in the following generalization of the noisy voter model.

\section{Scaled voter model and anomalous diffusion}

\label{sec:Voter-time-dependent-herding}

In this section, we will study the noisy voter model with the time-dependent
herding behavior intensity, $h(t)$. Here we assume that the herding behavior intensity $h$
in the SDE~(\ref{eq:sdex}) depends on the real time and the independent transition rates
are proportional to the herding behavior intensity
$r_{i}=\varepsilon_{i}h(t)$:
\begin{equation}
dx=h(t)\left[\varepsilon_{1}\left(1-x\right)-\varepsilon_{2}x\right]dt+\sqrt{2h(t)x\left(1-x\right)}dW_{t}\, .\label{eq:exact-Voter-all-h(t)}
\end{equation}

If we assume that the herding intensity is a power-law function of time
\begin{equation*}
h(t)= \gamma_{s}t^{\gamma_{s}-1},
\end{equation*} 
then SDE~(\ref{eq:exact-Voter-all-h(t)}) becomes
\begin{equation}
dx=\gamma_{s}t^{\gamma_{s}-1}\left[\varepsilon_{1}\left(1-x\right)-\varepsilon_{2}x\right]dt+\sqrt{2x\left(1-x\right)}\sqrt{\gamma_{s}}t^{\frac{\gamma_{s}-1}{2}}dW_{t}\, .\label{eq:Scaled-Voter}
\end{equation}
By performing a time scale change $t\rightarrow t_{s}=t^{\gamma_{s}}$,
by using relation $dt_{s}=\gamma_{s}t^{\gamma_{s}-1}dt$ and by using the definition
of SBM, SDE~(\ref{eq:SBM-in-real-time}), we can show that
\begin{equation}
dx=\left[\varepsilon_{1}\left(1-x\right)-\varepsilon_{2}x\right]dt_{s}+\sqrt{2x\left(1-x\right)}dW_{t_{s}}.
\end{equation}

The process described by SDE~(\ref{eq:Scaled-Voter}) in scaled time
$t_{s}$ is identical to the original noisy voter model, SDE~(\ref{eq:sdex}),
in real time $t$. Therefore from now on a process described by SDE~(\ref{eq:Scaled-Voter})
we will refer to as the scaled voter model.

For small $x$ values $(x\ll1)$, we can neglect higher $x$ members
in the diffusion term $\sqrt{2h(t)x\left(1-x\right)}\approx\sqrt{2h(t)x}$
\begin{equation}
dx=h(t)\left[\varepsilon_{1}\left(1-x\right)-\varepsilon_{2}x\right]dt+\sqrt{2h(t)}\sqrt{x}dW_{t}\, .\label{eq:Voter-small-X}
\end{equation}

Let us introduce the following notation:
\begin{equation}
\alpha(t)=(\varepsilon_{1}+\varepsilon_{2})h(t),\,\,\,\,\,\beta(t)=\varepsilon_{1}h(t),\,\,\,\,\,k(t)=\sqrt{2h(t)}.\label{eq:CIR-time-coeff}
\end{equation}

Then we can see that (for $x\ll1$) SDE~(\ref{eq:exact-Voter-all-h(t)})
can be well approximated by the CIR process \cite{Cox1985} with time-dependent
coefficients 
\begin{equation}
dx=-\left[\alpha(t)x-\beta(t)\right]dt+k(t)\sqrt{x}dW_{t}\, .\label{eq:SDE-CIR-time-depent-coeffs}
\end{equation}

In Refs. \cite{Masoliver2016,Araneda2020} it has been shown that if condition 
\begin{equation}
\frac{2\beta(t)}{k^{2}(t)}=\varepsilon_{1}=\mathrm{const},\label{eq:Solution-valibily-condition}
\end{equation}
is satisfied then time and space variables can be separated 
in the Fokker-Planck equation by using time transformation. 
Therefore the Fokker-Planck equation corresponding to Eq.~(\ref{eq:SDE-CIR-time-depent-coeffs}) 
can be solved by using a well-known solution  in the form of Bessel functions, 
then the transition probability $P(x,t|x_{0},0)$ 
of the CIR process with time-dependent coefficients is \cite{Masoliver2016}
\begin{equation}
P(x,t|x_{0},0)=\frac{1}{\phi(t)}\left(\frac{x}{x_{0}e^{-\tau(t)}}\right)^{(\varepsilon_{1}-1)/2}\exp\left(-\frac{x+x_{0}e^{-\tau(t)}}{\phi(t)}\right)I_{\varepsilon_{1}-1}\left(\frac{2}{\phi(t)}\sqrt{xx_{0}e^{-\tau(t)}}\right).\label{eq:CIR-TransProb}
\end{equation}
Here $x_{0}$ is the initial condition and we set the
initial time to $t_{0}=0$. For voter models with linear herding such as SDE~(\ref{eq:sdex}) coefficients $\beta(t)$ and $k(t)$ always appear in such form that the condition
defined by Eq.~(\ref{eq:Solution-valibily-condition}) are satisfied
for all $t$. The aforementioned condition ensures that the diffusion and drift coefficients only influence the relaxation of the process to the steady state. But the steady-state distribution of the stochastic process remains the same as if the coefficients would be constant.
An analytical expression for the transition
probability also can be found for $\beta(t)=0$ (if $h\neq0$,
$\varepsilon_{1}=0$) \cite{Masoliver2016}, however, such a solution
is not useful in the context of anomalous diffusion because the process
tends to the singularity at zero as time progresses \cite{Masoliver2016,Gan2015}.
Time-dependent functions $\phi(t)$ and $\tau(t)$ are the time integrals
of the CIR coefficients: 
\begin{align}
\tau(t) & =\int_{0}^{t}\alpha(s)ds,\label{eq:TAU-integral}\\
\phi(t) & =\frac{1}{2}\int_{0}^{t}k^{2}(t^{\prime})\exp\left(-\int_{t^{\prime}}^{t}\alpha(s)ds\right)dt^{\prime}.\label{eq:PHI-integral}
\end{align}
Using Eq.~(\ref{eq:CIR-TransProb}) we can calculate the time-dependent
average of the $\kappa$th power of $x$:
\begin{equation}
\langle x^{\kappa}(t,x_{0})\rangle=\int_{0}^{\infty}y^{\kappa}P(x,t|x_{0},0)dx.\label{eq:Moments-generator}
\end{equation}
By inserting Eq.~(\ref{eq:CIR-TransProb}) into Eq.~(\ref{eq:Moments-generator})
and setting $\kappa=1$ we obtain a general formula for the first moment
of $x(t)$:
\begin{equation}
\langle x(t,x_{0})\rangle=x_{0}e^{-\tau(t)}+\varepsilon_{1}\phi(t).\label{eq:mean-GENERAL}
\end{equation}
By inserting Eq.~(\ref{eq:CIR-TransProb}) into Eq.~(\ref{eq:Moments-generator})
and setting $\kappa=2$ we obtain a general formula for the second moment
of $x(t)$:
\begin{equation}
\langle x^{2}(t,x_{0})\rangle=x_{0}^{2}e^{-2\tau(t)}+2x_{0}(1+\varepsilon_{1})e^{-\tau(t)}\phi(t)+\varepsilon_{1}(1+\varepsilon_{1})\phi^{2}(t).\label{eq:Second-moment-GEN}
\end{equation}
From Eqs.~(\ref{eq:mean-GENERAL}) and (\ref{eq:Second-moment-GEN}) it
follows
that the variance of $x(t)$ is
\begin{equation}
\mathrm{Var}[x(t|x_{0})]=\langle x^{2}(t,x_{0})\rangle-\langle x(t,x_{0})\rangle^{2}=\phi(t)\left(2e^{-\tau(t)}x_{0}+\varepsilon_{1}\phi(t)\right).\label{eq:variance-GENERAL}
\end{equation}

From now on, let us consider the case of power-law temporal scaling
of the herding behavior intensity function, $h(t)=\gamma_{s}t^{\gamma_{s}-1}$. Such
form of the herding behavior intensity function was chosen to introduce SBM 
(see SDE~(\ref{eq:SBM-in-real-time})) into the
ABM described by the SDE~(\ref{eq:exact-Voter-all-h(t)}). Without loss
of generality, let us set the diffusion coefficient to unity, $D=1$, in SDE~(\ref{eq:SBM-in-real-time}).
If the herding behavior intensity has such power-law temporal scaling form, then
from Eqs.~(\ref{eq:CIR-time-coeff}), (\ref{eq:TAU-integral}), and
(\ref{eq:PHI-integral}) it follows that
\begin{equation}
\phi(t)=\frac{1-e^{-(\varepsilon_{1}+\varepsilon_{2})t^{\gamma_{s}}}}{\varepsilon_{1}+\varepsilon_{2}},\label{eq:phi-pow-LAW}
\end{equation}
and
\begin{equation}
\tau(t)=(\varepsilon_{1}+\varepsilon_{2})t^{\gamma_{s}}.\label{eq:TAU-pow-LAW}
\end{equation}
Furthermore from Eqs.~(\ref{eq:mean-GENERAL}), (\ref{eq:phi-pow-LAW}),
and (\ref{eq:TAU-pow-LAW}) it follows that the time evolution of
the mean is 
\begin{equation}
\langle x(t,x_{0})\rangle=x_{0}e^{-at^{\gamma_{s}}}+b\left(1-e^{-at^{\gamma_{s}}}\right),\label{eq:mean-power-law}
\end{equation}
with
\begin{equation}
a=\varepsilon_{1}+\varepsilon_{2},\qquad b=\frac{\varepsilon_{1}}{\varepsilon_{1}+\varepsilon_{2}}.\label{eq:_a_b_CIR_parameters}
\end{equation}
In the context of the noisy voter model, $\varepsilon_{1}$ and $\varepsilon_{2}$
are the independent transition rates and we assume that they are
positive real numbers, therefore $a$ and $b$ are also positive real
numbers. Note that if we set $\gamma_{s}=1$, Eq.~(\ref{eq:mean-power-law})
reduces to the mean formula for the CIR process with constant coefficients
\cite{Cox1985}.

Let us consider the time evolution of the mean $\langle x(t,x_{0})\rangle$,
Eq.~(\ref{eq:mean-power-law}). In the case when the initial
position $x_{0}$ is set such as $x_{0}\ll b$, by performing the Taylor series
expansion we can show that the time evolution
of the mean exhibits power-law scaling for intermediate times
\begin{equation}
\langle x(t,x_{0})\rangle=\varepsilon_{1}t^{\gamma_{s}},\quad t_{x_{0}}<t<t_{c}\, .\label{eq:mean-ANOMALOUS}
\end{equation}
Here 
\begin{equation}
t_{x_{0}}=\bigg|\frac{x_{0}}{a(b-x_{0})}\bigg|^{\frac{1}{\gamma_{s}}}=\bigg|\frac{x_{0}}{\varepsilon_{1}-x_{0}(\varepsilon_{1}+\varepsilon_{2})}\bigg|^{\frac{1}{\gamma_{s}}},
\end{equation}
is the time moment at which the influence of initial position $x_{0}$
is forgotten. If we set the initial value $x_{0}=0$ then $t_{x_{0}}=0$
this means that in this case, power-law scaling should start instantly.
 $t_{c}$ defines the critical time value at which power-law scaling of the mean
 stops:
\begin{equation}
t_{c}=\frac{1}{a^{\frac{1}{\gamma_{s}}}}=\frac{1}{\big(\varepsilon_{1}+\varepsilon_{2}\big)^{\frac{1}{\gamma_{s}}}}.
\end{equation}
For larger times $t>t_{c}$, power-law scaling of the mean subsides
and the mean starts to tend to its steady-state value.

If the herding behavior intensity is a power-law function of time, 
$h(t)= \gamma_{s}t^{\gamma_{s}-1}$, then from
Eqs.~(\ref{eq:variance-GENERAL}), (\ref{eq:phi-pow-LAW}), and (\ref{eq:TAU-pow-LAW})
it follows that the variance of $x(t)$ is
\begin{equation}
\mathrm{Var}[x(t|x_{0})]=\langle x^{2}(t,x_{0})\rangle-\langle x(t,x_{0})\rangle^{2}=\frac{2x_{0}}{a}\left(e^{-at^{\gamma_{s}}}-e^{-2at^{\gamma_{s}}}\right)+\frac{b}{a}\left(1-e^{-at^{\gamma_{s}}}\right)^{2}.\label{eq:variance-power-LAW}
\end{equation}
Parameters $x_{0}$, $a$, and $b$ are the same as for the mean $\langle x(t,x_{0})\rangle$
defined by Eq.~(\ref{eq:mean-power-law}). For parameter $\gamma_{s}=1$
variance $\mathrm{Var}[x(t|x_{0})]$ reduces to a variance formula for the CIR process
with constant parameters \cite{Cox1985} (see Eq. (19) with $\sigma=\sqrt{2}$
in Ref. \cite{Cox1985}).

Now let us consider the time evolution of the variance $\mathrm{Var}[x(t|x_{0})]$,
given by Eq.~(\ref{eq:variance-power-LAW}) for various $x$ initial
values $x_{0}$. In the case when the initial position $x_{0}=0$,
the variance takes the form 
\begin{equation}
\mathrm{Var}[x(t|0)]=\frac{b}{a}\left(1-e^{-at^{\gamma_{s}}}\right)^{2}=\frac{b}{a}\left(1-2e^{-at^{\gamma_{s}}}+e^{-2at^{\gamma_{s}}}\right).
\end{equation}

After performing the Taylor series expansion
we see that for $x_{0}=0$ and times $t<t_{c}$, the variance exhibits power-law scaling
\begin{equation}
\mathrm{Var}[x(t|0)]=\varepsilon_{1}t^{2\gamma_{s}},\quad t<t_{c}=1/a^{\frac{1}{\gamma_{s}}}.\label{eq:Double-p-law_Exponent}
\end{equation}

In this case, when diffusion starts at zero $x_{0}=0$ the variance
$\mathrm{Var}[x(t|0)]=\langle x^{2}(t,0)\rangle-\langle x(t,0)\rangle^{2}$
behavior is the same as MSD, this is
not the case for other initial positions $x_{0}$. Therefore, from
Eq.~(\ref{eq:Double-p-law_Exponent}), it follows that for times smaller
than $t_{c}$, the SDE~(\ref{eq:Voter-small-X}) with power-law herding, $h(t)$,
generated signal MSD scales with a double exponent compared to
standard SBM $\langle x^{2}(t)\rangle\sim t^{\gamma_{s}}$. Therefore
we can observe anomalous diffusion for $0<\gamma_{s}<1$, and for
$\gamma_{s}>1$, we can even observe superballistic motion \cite{Metzler2000}.

Some authors use only MSD as an indicator of the anomalous
diffusion \cite{Metzler2000,Metzler2000JCHempPhys,Metzler2004,Cherstvy2014}.
However, we decided to use the variance as an anomalous diffusion
indicator instead of MSD, because variance takes into consideration
the influence of initial value $x_{0}$. As we will see later the
introduction of initial position $x_{0}$ can lead to more interesting
results.

In general, the evolution of variance for times $t<t_{c}$
can be expressed as 
\begin{equation}
\mathrm{Var}[x(t|x_{0})]=2x_{0}t^{\gamma_{s}}+3a\left(\frac{b}{3}-x_{0}\right)t^{2\gamma_{s}},\quad t<t_{c}=1/a^{\frac{1}{\gamma_{s}}}.\label{eq:Anomalous-any-x-zero}
\end{equation}

For $x_{0}>b/3$, the second term (the one proportional to $t^{2\gamma_{s}}$) in Eq.~(\ref{eq:Anomalous-any-x-zero})  becomes negative and then
the first term dominates for all times up to $t_{c}$. Therefore for
the initial position $x_{0}>b/3$, the SDE~(\ref{eq:Voter-small-X}) generated
signal should exhibit the anomalous diffusion
\begin{equation}
\mathrm{Var}[x(t|x_{0}>b/3)]=2x_{0}t^{\gamma_{s}},\quad t<t_{c}=1/a^{\frac{1}{\gamma_{s}}}.
\end{equation}

For the case with $x_{0}<b/3$, one no longer can ignore the second
term in Eq.~(\ref{eq:Anomalous-any-x-zero}). Consequently, we should observe
double power-law scaling in variance
\begin{equation}
\mathrm{Var}[x(t|x_{0}<b/3)]=\begin{cases}
2x_{0}t^{\gamma_{s}}, & 0<t<t_{b}\ ,\\
3a\left(\frac{b}{3}-x_{0}\right)t^{2\gamma_{s}}, & t_{b}<t<t_{c}\ .
\end{cases}\label{eq:Double power-law}
\end{equation}

\begin{equation}
t_{b}=\left(\frac{2x_{0}}{3a(\frac{b}{3}-x_{0})}\right)^{1/\gamma_{s}},
\end{equation}
Here $t_{b}$ is the time moment when more ballistic diffusion starts (for $\gamma_{s}>1$).

\begin{figure}
\centering
\includegraphics[width=0.4\textwidth]{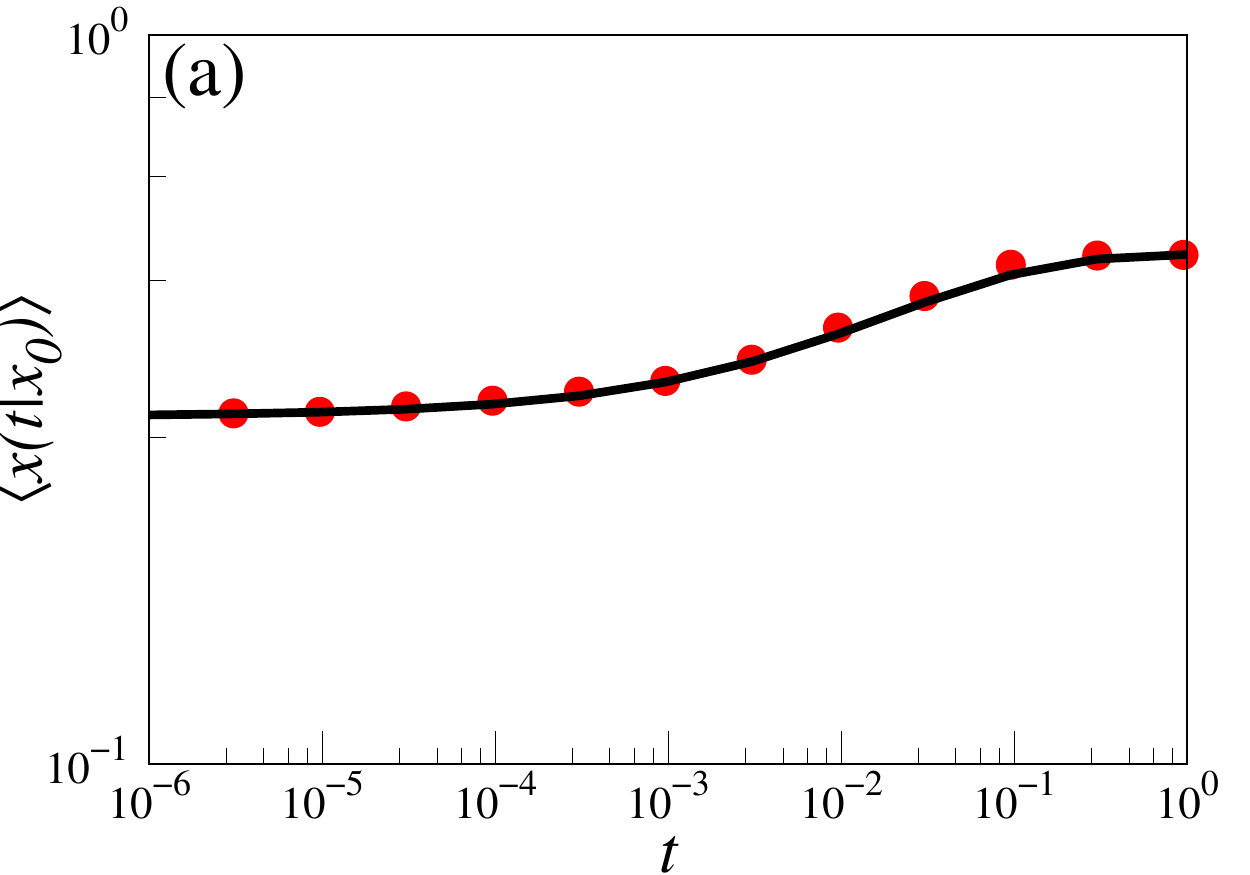}\hspace{0.05\textwidth}\includegraphics[width=0.4\textwidth]{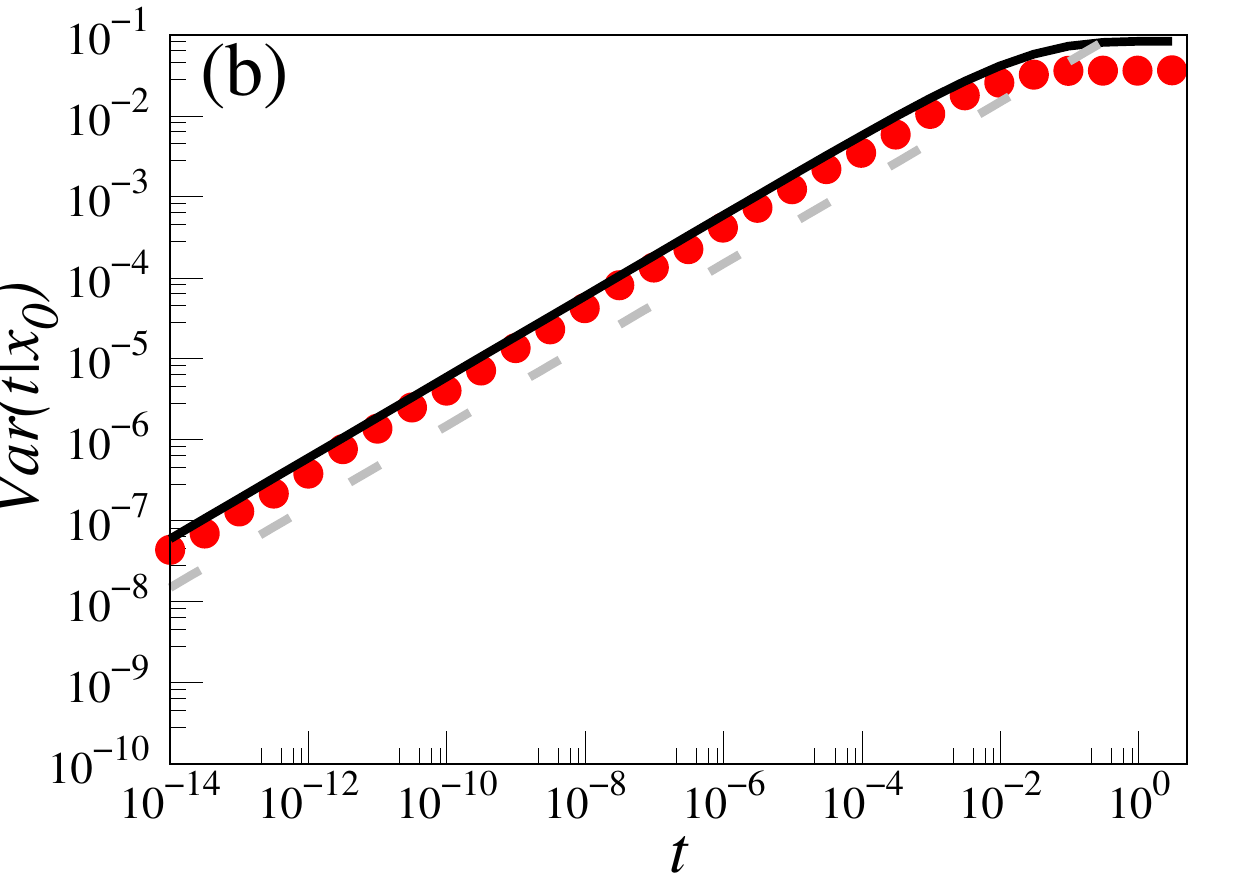}\\\includegraphics[width=0.4\textwidth]{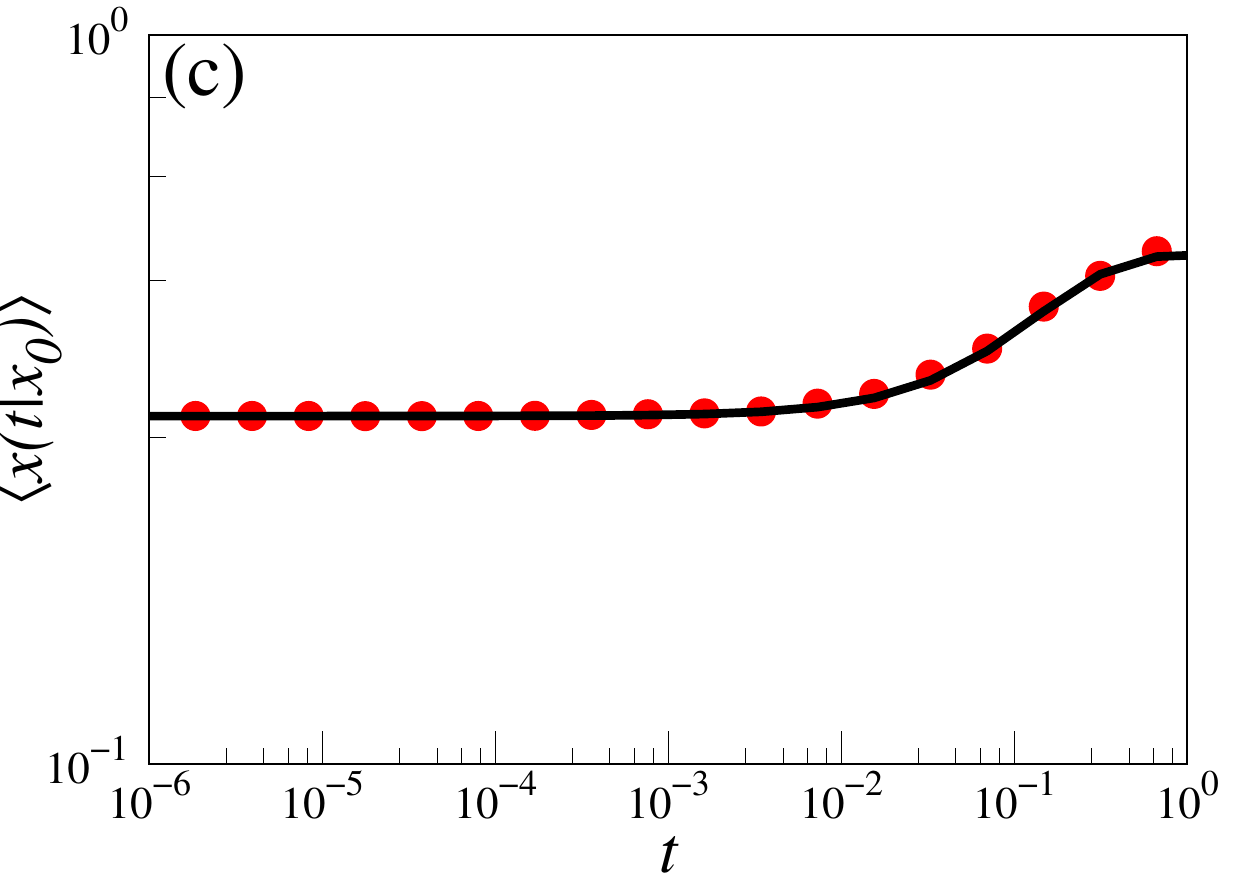}\hspace{0.05\textwidth}\includegraphics[width=0.4\textwidth]{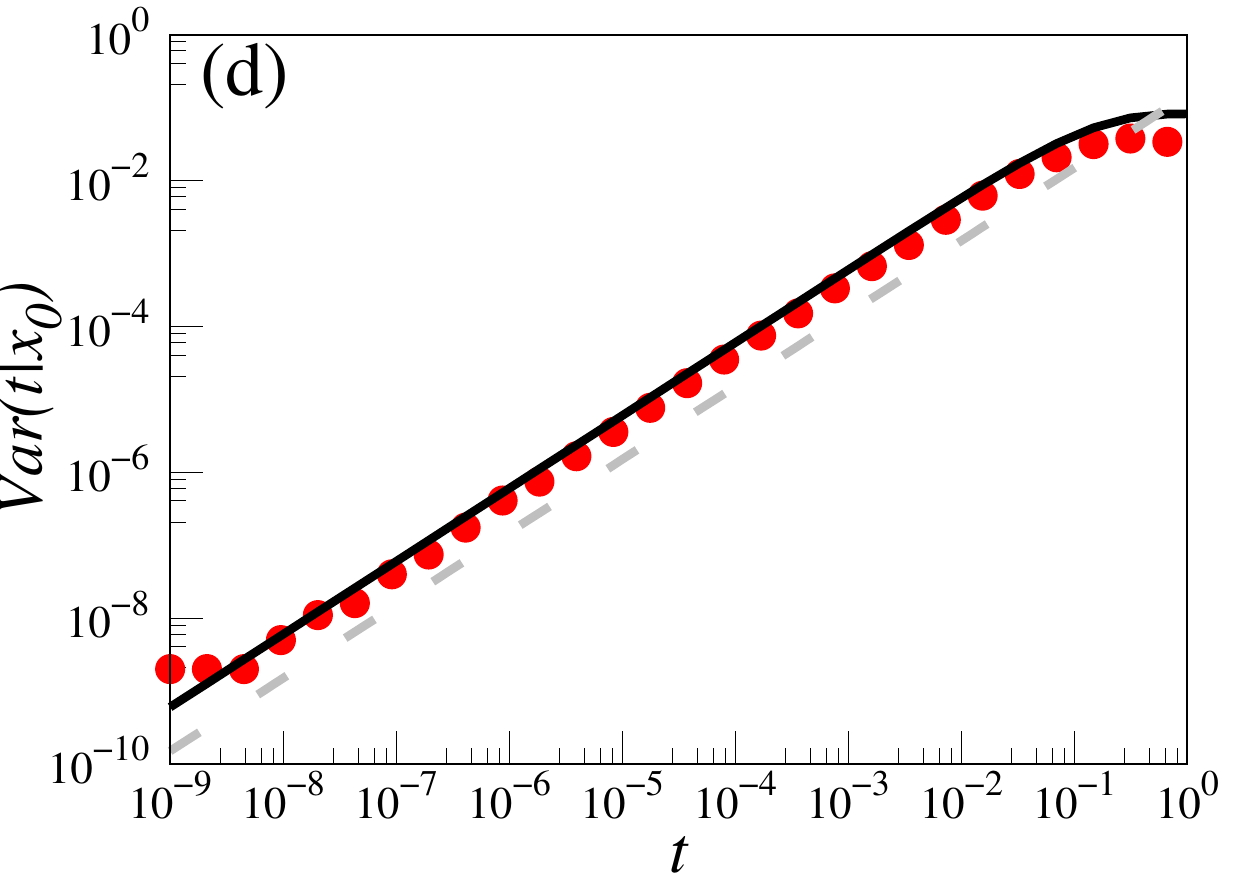}\\\includegraphics[width=0.4\textwidth]{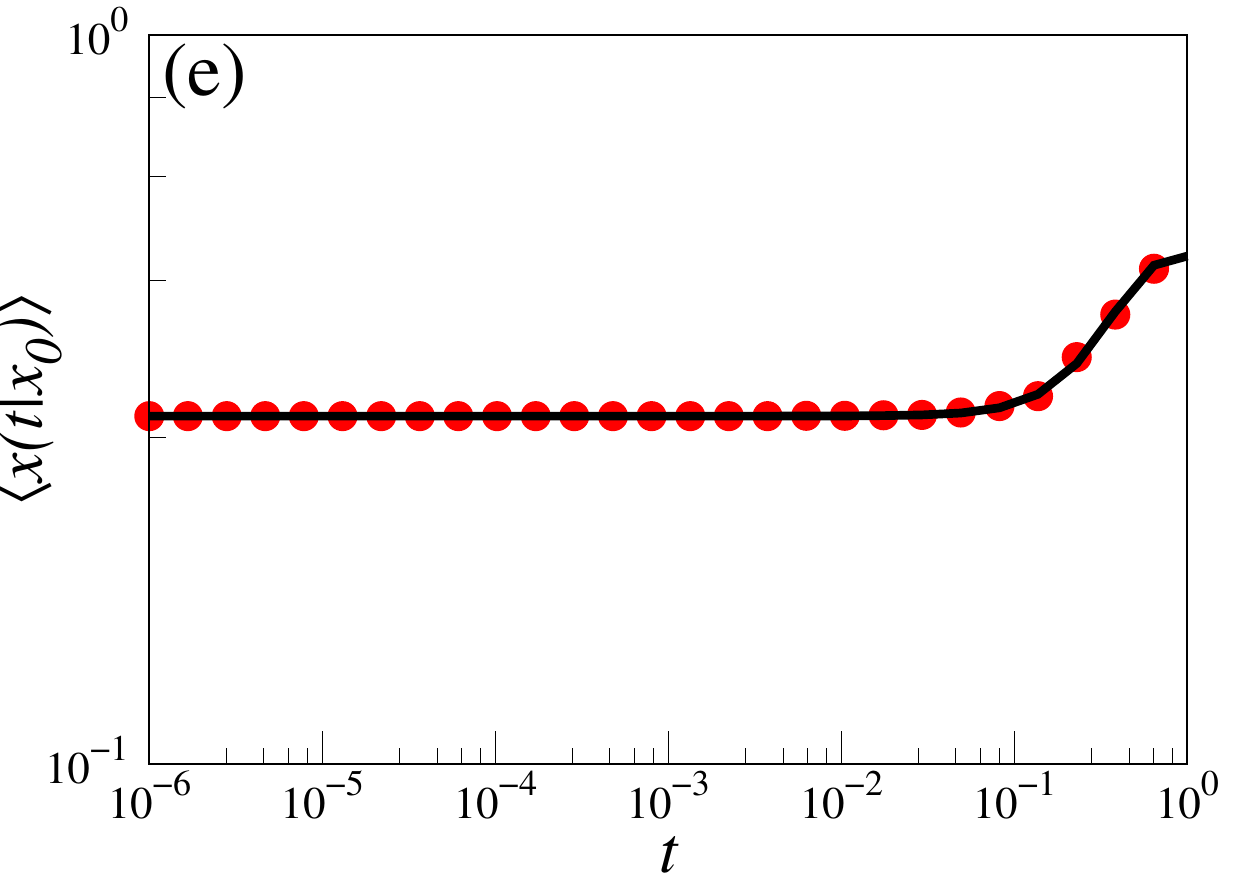}\hspace{0.05\textwidth}\includegraphics[width=0.4\textwidth]{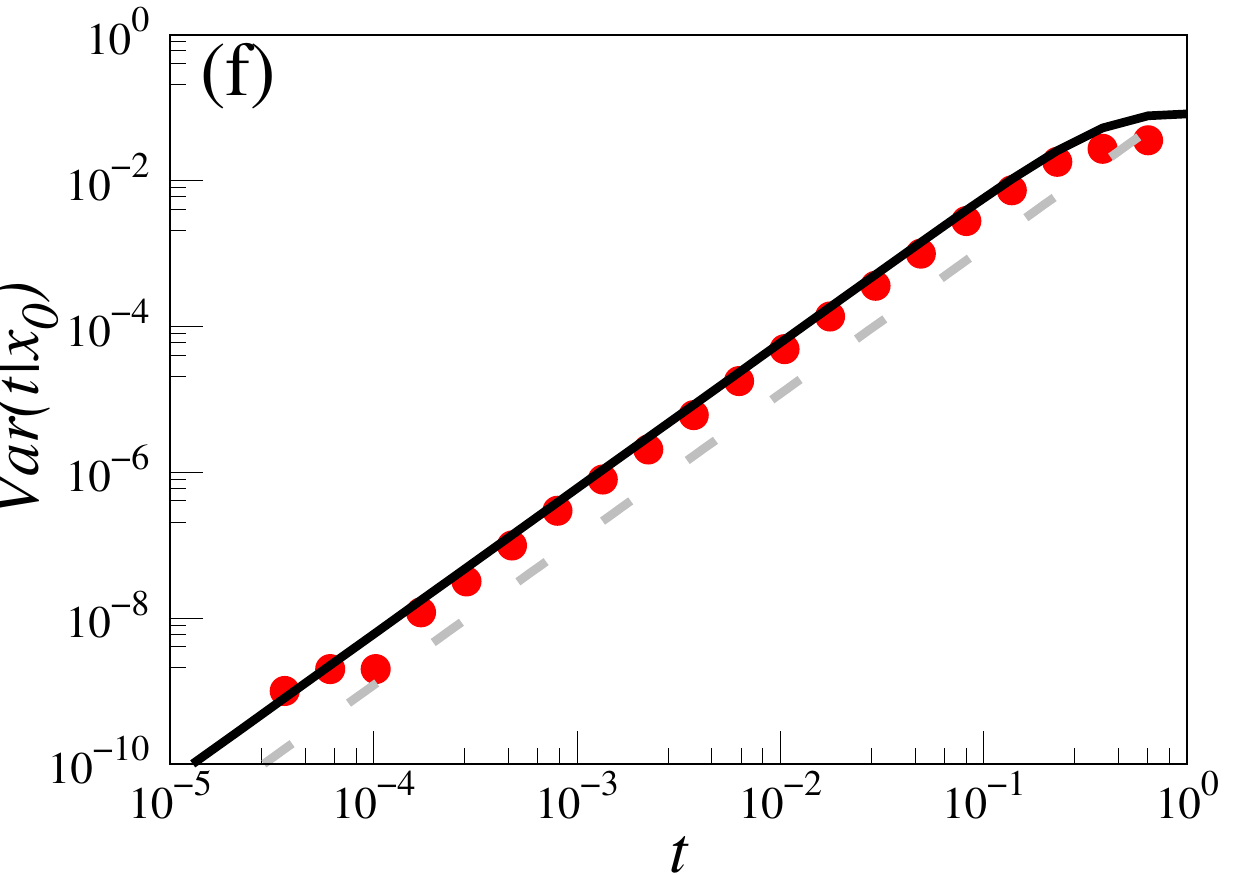}
\caption{Temporal evolution of the mean and the variance in the scaled voter model,
SDE~(\ref{eq:Scaled-Voter}), for various
parameter $\gamma_{s}$ values. Red points represent the results of numerical simulations.
Black (solid) lines are calculated using analytical Eq.~(\ref{eq:variance-power-LAW}),
and grey dashed lines show the power-law dependence on time $\sim t^{\gamma_{s}}$.
The common parameter values were set as follows: $x_{0}=0.3$ and
$\varepsilon_{1}=\varepsilon_{2}=3.0$ ($b/3=1/6$). SBM anomalous diffusion
exponent is different for the three cases shown: $\gamma_{s}=1/2$ for (a)
and (b), $\gamma_{s}=1$ for (c) and (d), $\gamma_{s}=2$ for (e) and (e).}
 \label{fig:mean-var-eps3-x0-03 gg b3}
\end{figure}

\begin{figure}
\centering
\includegraphics[width=0.4\textwidth]{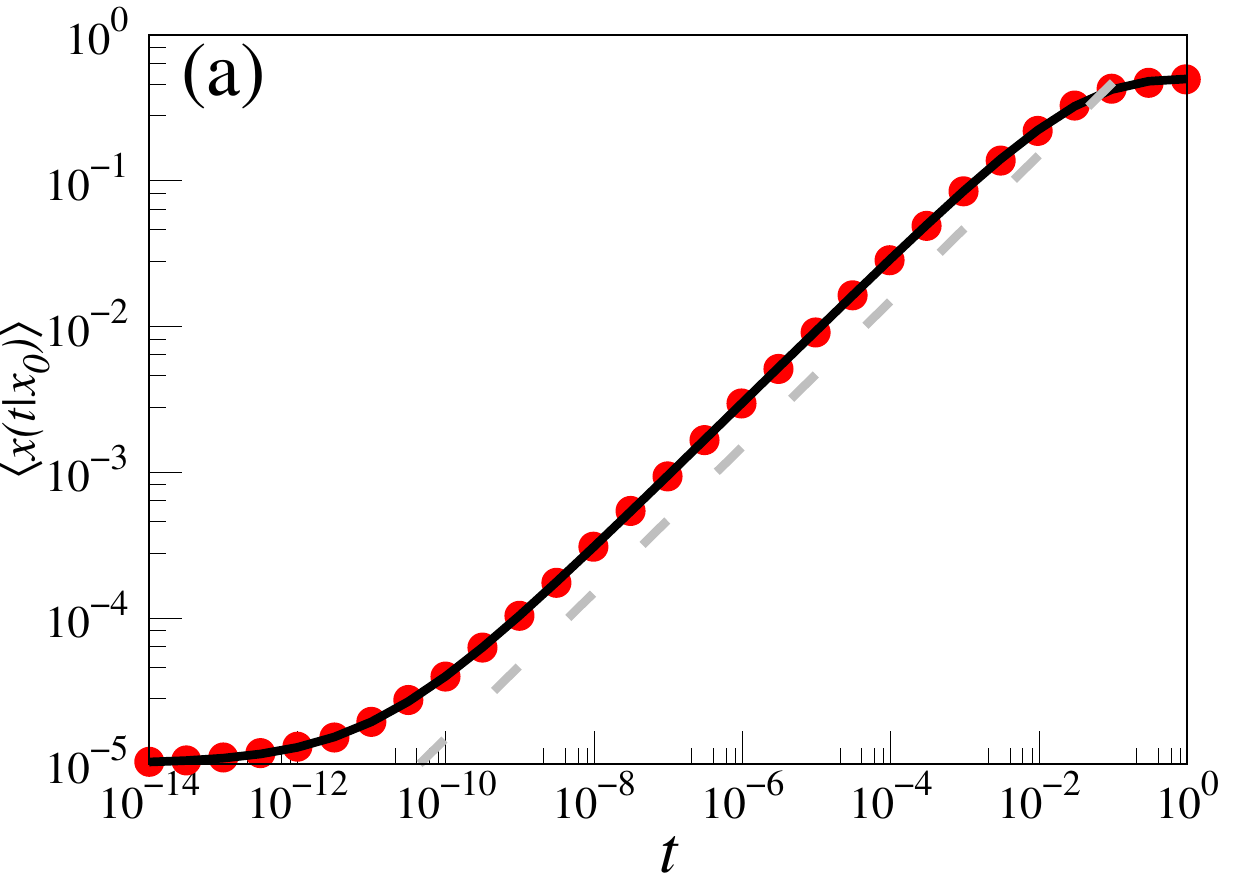}\hspace{0.05\textwidth}\includegraphics[width=0.4\textwidth]{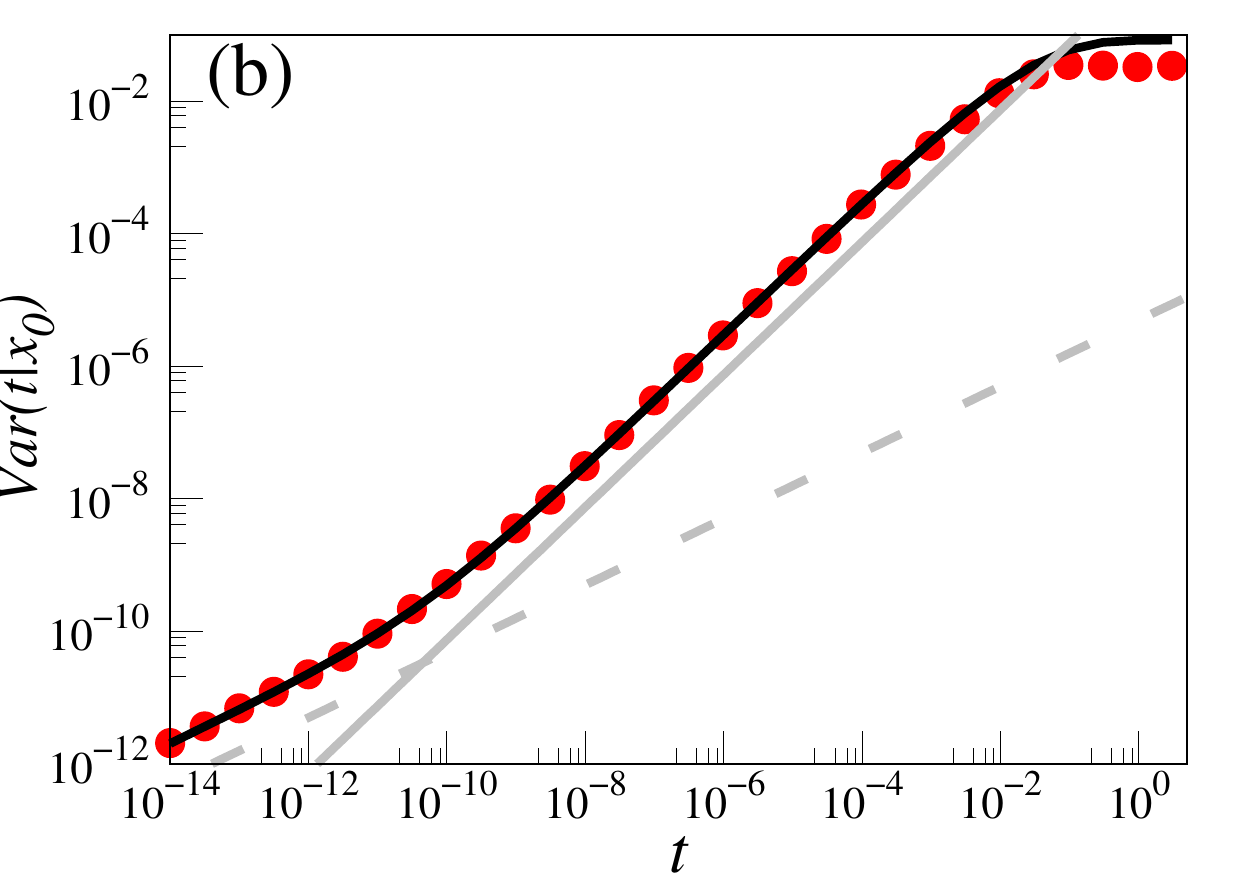}\\\includegraphics[width=0.4\textwidth]{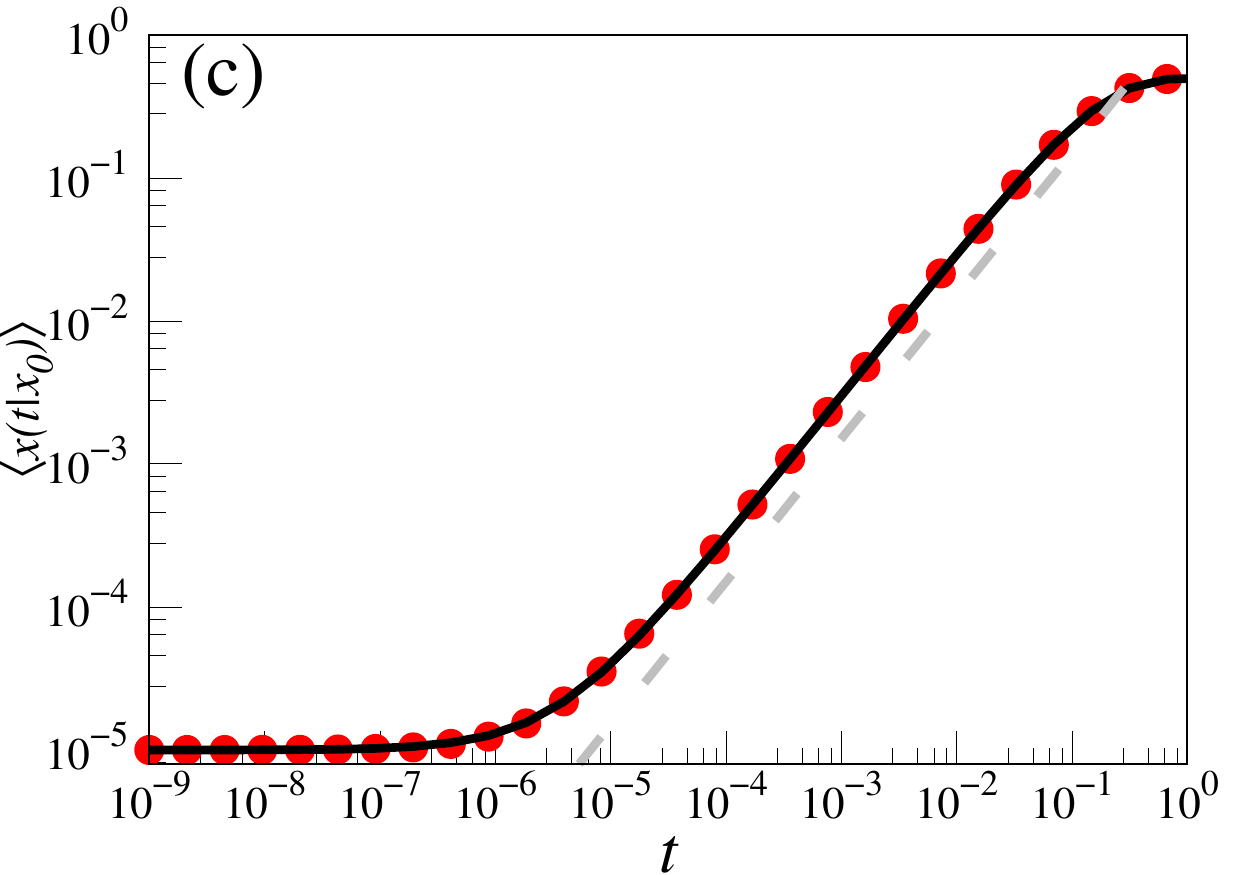}\hspace{0.05\textwidth}\includegraphics[width=0.4\textwidth]{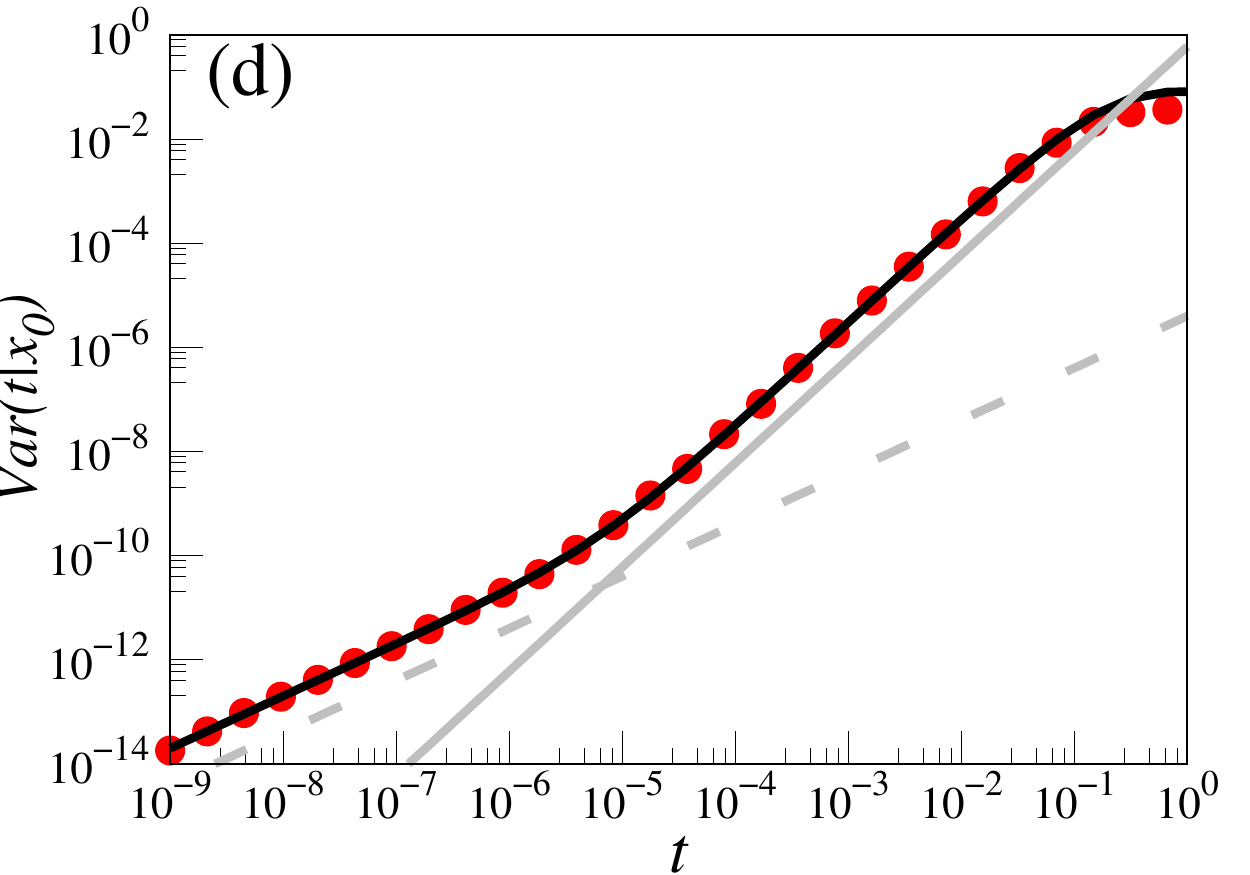}\\\includegraphics[width=0.4\textwidth]{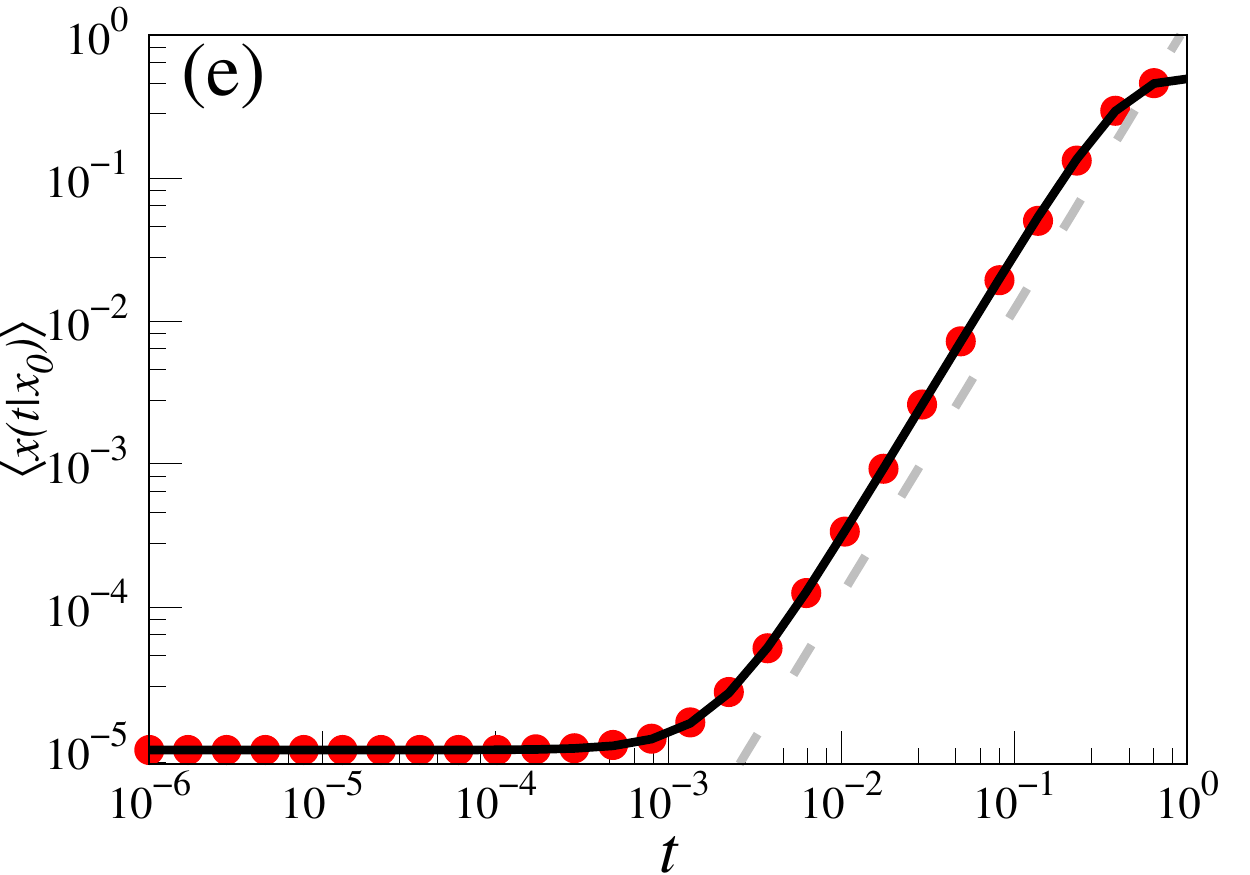}\hspace{0.05\textwidth}\includegraphics[width=0.4\textwidth]{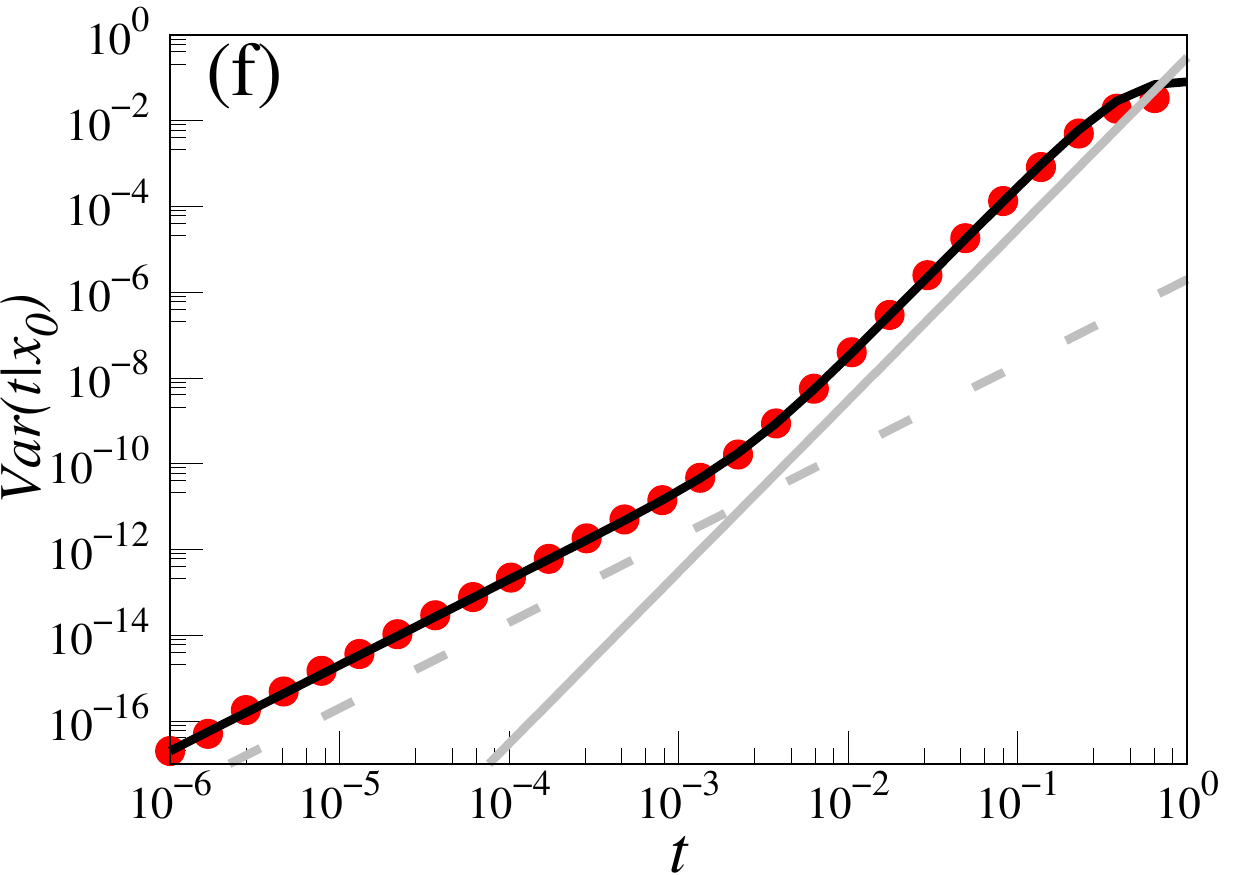}
\caption{Temporal evolution of the mean and the variance in the scaled voter model,
SDE~(\ref{eq:Scaled-Voter}), for various parameter $\gamma_{s}$ values. 
Red points represent the results of numerical simulations.
Black (solid) lines are calculated
using analytical Eq.~(\ref{eq:variance-power-LAW}), and grey lines
show the power-law dependence on time $\sim t^{\gamma_{s}}$ (dashed)
and $\sim t^{2\gamma_{s}}$ (solid) respectively.
The common parameter values were set as follows: 
$x_{0}=10^{-5}$ and $\varepsilon_{1}=\varepsilon_{2}=3.0$ ($b/3=1/6$) for all
pictures. SBM anomalous diffusion exponent is different for the three cases
shown: $\gamma_{s}=1/2$ for (a) and (b), $\gamma_{s}=1$ for (c) and (d), $\gamma_{s}=2$ for (e) and (e).}
 \label{fig:means-veps3-double-p-law}
\end{figure}

The mean and variance power-law scaling for the scaled voter model is very sensitive
to the initial position $x_{0}$. In the case of $x_{0}>b/3$
(here $b=\frac{\varepsilon_{1}}{\varepsilon_{1}+\varepsilon_{2}}$
and $\varepsilon_{i}$ are transition rates) variance exhibits the
same anomalous power-law scaling
as SBM up to critical time $t_{c}$ (see Fig.~\ref{fig:mean-var-eps3-x0-03 gg b3}).
After critical time $t_{c}$ moments tends to their steady-state values.
This is quite an unexpected result because other types of noisy voter
model transformation lead to inverse power-law decay from the initial
position to steady-state values for bought mean and variance for large
initial value \cite{Kazakevicius2021PRE}. In the case
of $x_{0}<b/3$ (see Fig.~\ref{fig:means-veps3-double-p-law}) we
can observe double power-law scaling of variance (see Eq~(\ref{eq:Double power-law})).
Until the influence of initial position $x_{0}$ is forgotten the variance
exhibits the same anomalous scaling as SBM up to time $t_{b}$. After time $t_{b}$ the variance
starts growing with the doubled exponent.
For $0.5<\gamma_{s}<1$, we can observe both types of
anomalous diffusion: subdiffusion transitioning into superdiffusion.
For $1<\gamma_{s}<2$, the superdiffusion transitioning into 
superballistic motion \cite{Metzler2000}.
For $0<\gamma_{s}<1$, such double power-law scaling has only
been obtained in more complex models such as the Galilei variant time-fractional
diffusion-advection equations \cite{Metzler2000,Metzler2000JCHempPhys}.
Numerical simulation shows that carrier-based transport through a line
of cells exhibits such double power-law scaling
with anomalous diffusion exponent $\gamma=0.59$
\cite{Kruse2008}. Continuous time random walk models suggest that
inverse double power-law scaling the of mean current can occur in amorphous
semiconductors \cite{Barkai2001}. Therefore, the ability to reproduce
double power-law scaling
of the variance might make our model more applicable
to a variety of physical and sociophysical systems.

\section{First passage time distribution of the scaled voter model}

\label{sec:FPTD}

In this section, we will obtain the FPTD for the special case of the noisy
voter model described by SDE~\eqref{eq:exact-Voter-all-h(t)} when
parameters $\varepsilon_{1}=\varepsilon_{2}=\varepsilon=1/2$.
In addition, we discuss how the scaled voter model can be differentiated from
other long-range memory processes using variance time dependence and obtained FPTD.

The It\^o SDE~\eqref{eq:exact-Voter-all-h(t)} can be transformed into SDE
with additive noise by introducing  new variable $y(x)=\sqrt{2}\arcsin(\sqrt{1-x})$. By using It\^o's formula \cite{Gardiner2004} we obtain SDE for a new variable $y(x(t))$
\begin{equation}
dy=h(t)F(y)dt-\sqrt{h(t)}dW_{t}\,.
\end{equation}

The drift term time-independent part $F(y)$ is 
\begin{equation}
F(y)=\frac{\varepsilon_{2}-\varepsilon_{1}}{\sqrt{2}}\frac{1}{\sin\big(\sqrt{2}y\big)}+\frac{\varepsilon_{1}+\varepsilon_{2}-1}{\sqrt{2}}\cot\big(\sqrt{2}y\big).\label{eq:for-FPT-y}
\end{equation}
Because the $x(t)$ process is bounded in interval $[0,1]$ its transformation
$y(x(t))$ is bounded in $[0,\pi/\sqrt{2}]$ due to force $F(y)$.

If the condition is satisfied $\varepsilon_{1}=\varepsilon_{2}=\varepsilon$
the $F(y)$ takes a simpler form and the SDE~\eqref{eq:for-FPT-y} becomes
\begin{equation}
dy=h(t)\frac{2\varepsilon-1}{\sqrt{2}}\cot\big(\sqrt{2}y\big)dt+\sqrt{h(t)}dW_{t}\,.\label{eq:for-FPT-y-epsilon}
\end{equation}

The minus sign can be dismissed because $dW$ is statistically equivalent
to $-dW$. In the case of time-independent herding behavior intensity SDE~\eqref{eq:for-FPT-y-epsilon}
has been obtained by using different nonlinear transformations \cite{Gontis2020PhysA}.

In the special case when $\varepsilon=1/2$, there is no drift force
and the process $y(t)$ can be described by a simpler SDE: 
\begin{equation}
dy=\sqrt{h(t)}dW_{t}\,.
\end{equation}

As in the previous section, we set the time-dependent herding behavior intensity
to be a power-law function of time $h(t)=\gamma_{s}t^{\gamma_{s}-1}$.
In this case, the process $y(t)$ is a special case of SBM with a diffusion
coefficient equal to $D=1/2$:
\begin{equation}
dy=t^{\frac{\gamma_{s}-1}{2}}\sqrt{\gamma_{s}}dW_{t}\,.
\end{equation}

In this special case, the scaled voter model can be interpreted
as a nonlinear transformation of SBM. Because $y$ is SBM, therefore
its FPTD according to Eq.~(\ref{eq:FPTD-SBM}) is
\begin{equation}
f_{y}(T)=\frac{|y_{0}-a_{y}|\gamma_{s}}{\sqrt{2\pi}}\frac{1}{T^{\gamma_{s}/2+1}}\exp\Bigg(-\frac{(y_{0}-a_{y})^{2}}{2T^{\gamma_{s}}}\Bigg).\label{eq:eq:FPTD-x-e1 and 2-1/2-Y}
\end{equation}
Here $|y_{0}-a_{y}|$ is the absolute value of the difference between
initial position $y_{0}$ and threshold $a_{y}$ (absorbing boundary). In
Eq.~(\ref{eq:eq:FPTD-x-e1 and 2-1/2-Y}), an exponential term can be
written in the form of $\exp(-(z/T^{-1}_{\Delta})^{\gamma_{s}})$, where $z=1/T$.
Therefore for short times ($z\gg T^{-1}_{\Delta}$), we have an exponential cut-off
 for the short passage times. For longer passage times FPTD decays as a power-law function
\begin{equation}
f_{y}(T)=\frac{|y_{0}-a_{y}|\gamma_{s}}{\sqrt{2\pi}}\frac{1}{T^{\gamma_{s}/2+1}},\quad T>T_{\Delta}=2^{-\frac{1}{\gamma_{s}}}|y_{0}-a_{y}|^{\frac{2}{\gamma_{s}}}.
\end{equation}

By remembering relation $y(x)=\sqrt{2}\arcsin(\sqrt{1-x})$ we obtain
FPTD for the scaled voter model $x$ (for parameters $\varepsilon_{1}=\varepsilon_{2}=1/2$)
\begin{equation}
f_{x}(T)=\frac{\gamma_{s}\Delta_{x}}{\sqrt{2\pi}}\frac{1}{T^{\gamma_{s}/2+1}}\exp\Bigg(-\frac{\Delta_{x}^{2}}{2T^{\gamma_{s}}}\Bigg).\label{eq:FPTD-x-e1 and 2-thesame and one half}
\end{equation}
Here $\Delta_{x}=\sqrt{2}\bigg|\arcsin\Big(\sqrt{1-x_{0}}\Big)-\arcsin\Big(\sqrt{1-a_{x}}\Big)\bigg|$
and $a_{x}$ absorption point in $x$ space.

\begin{figure}
\centering
\includegraphics[width=0.4\textwidth]{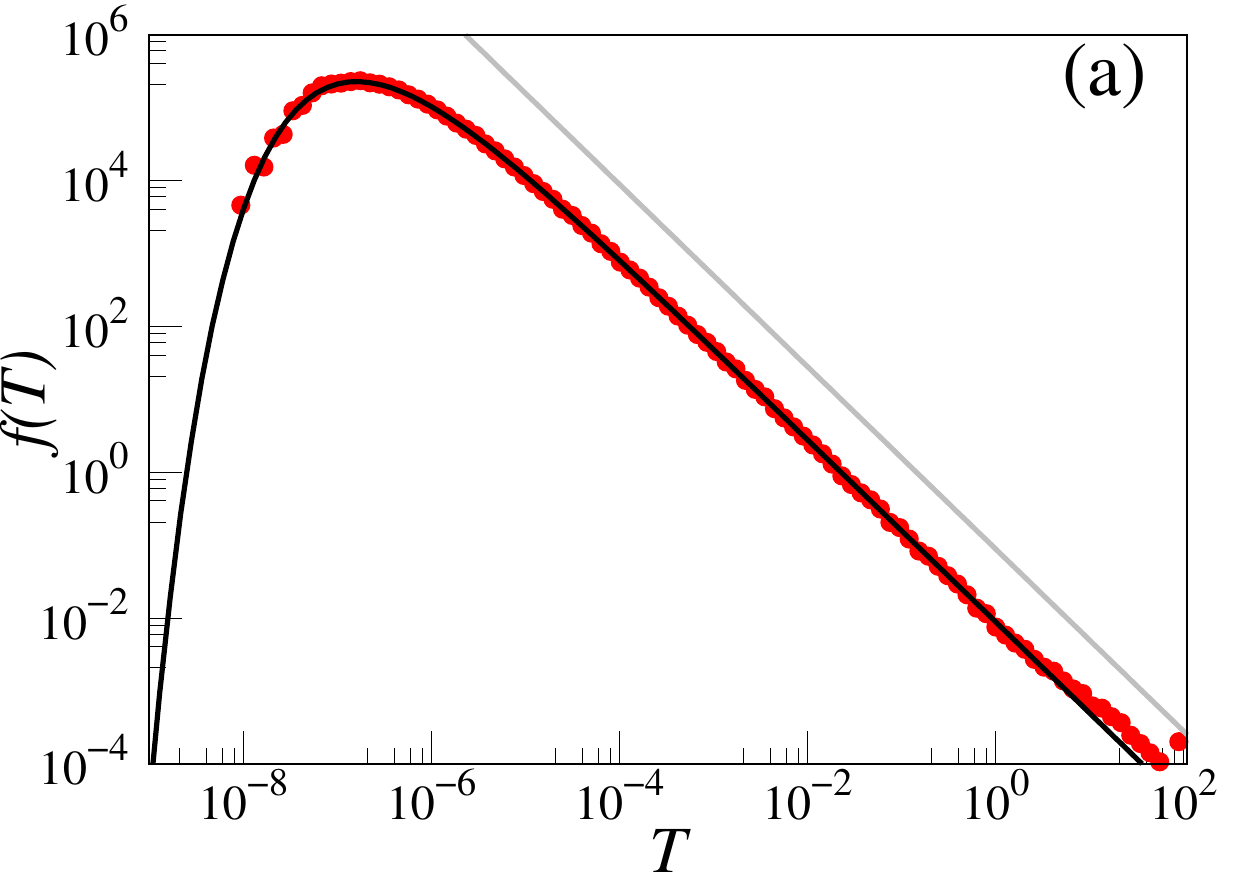}\hspace{0.05\textwidth}\includegraphics[width=0.4\textwidth]{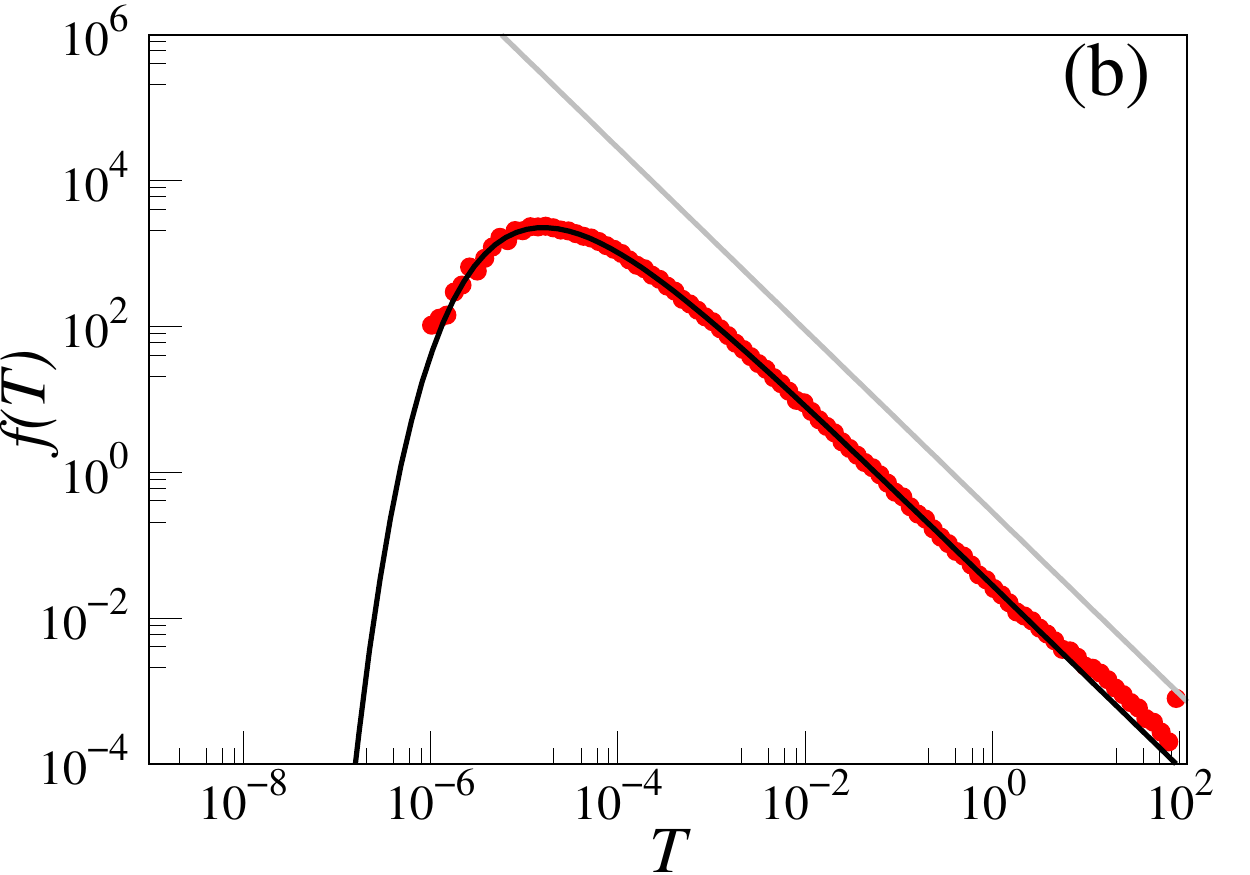}\\\includegraphics[width=0.4\textwidth]{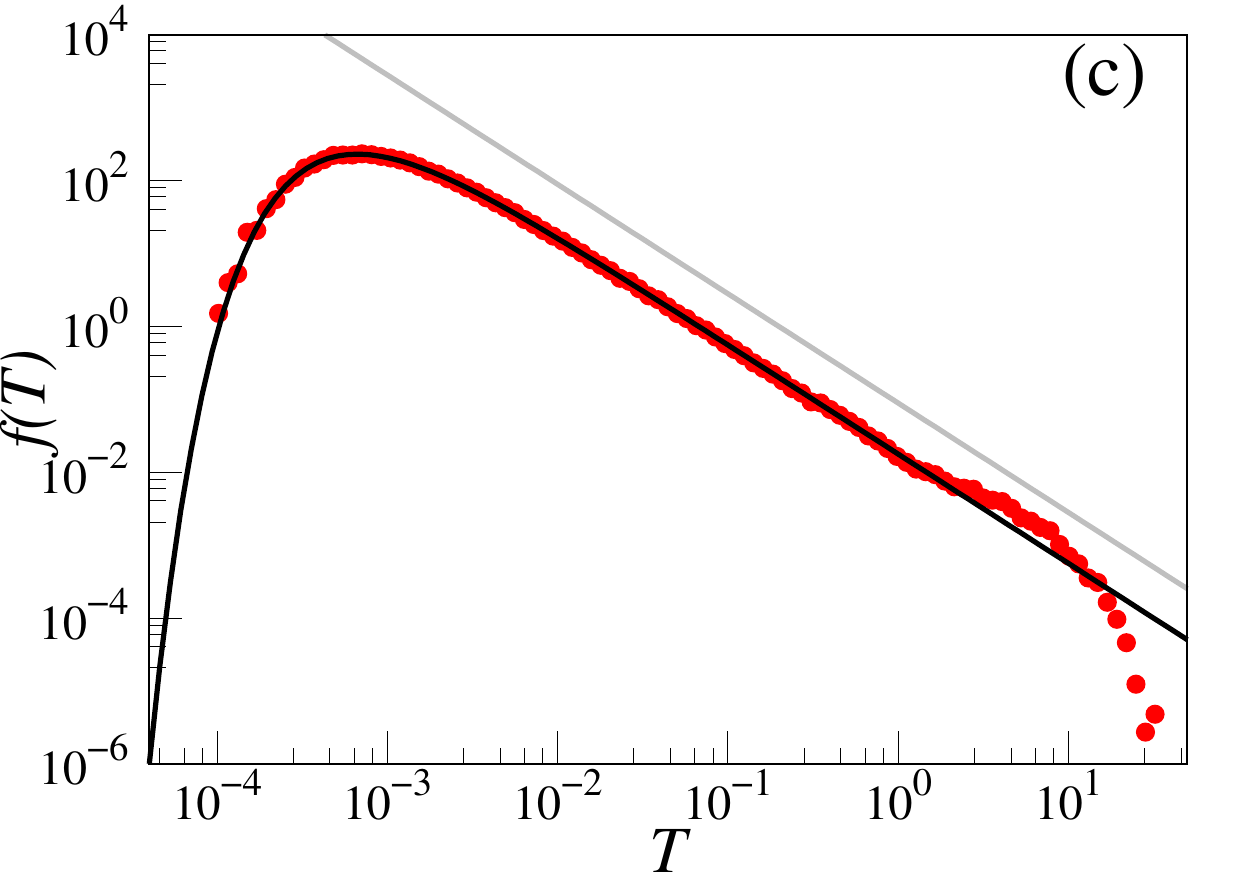}\hspace{0.05\textwidth}\includegraphics[width=0.4\textwidth]{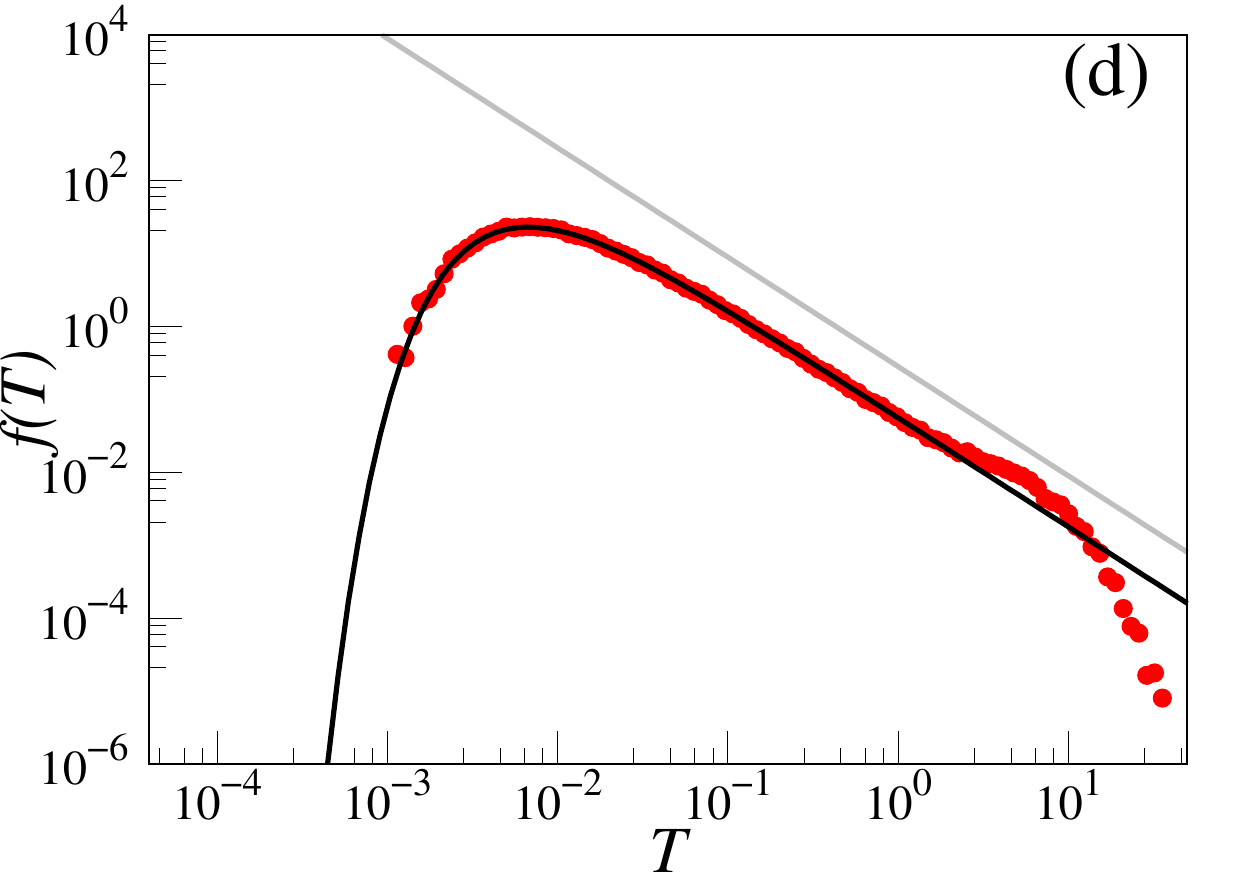}\\\includegraphics[width=0.4\textwidth]{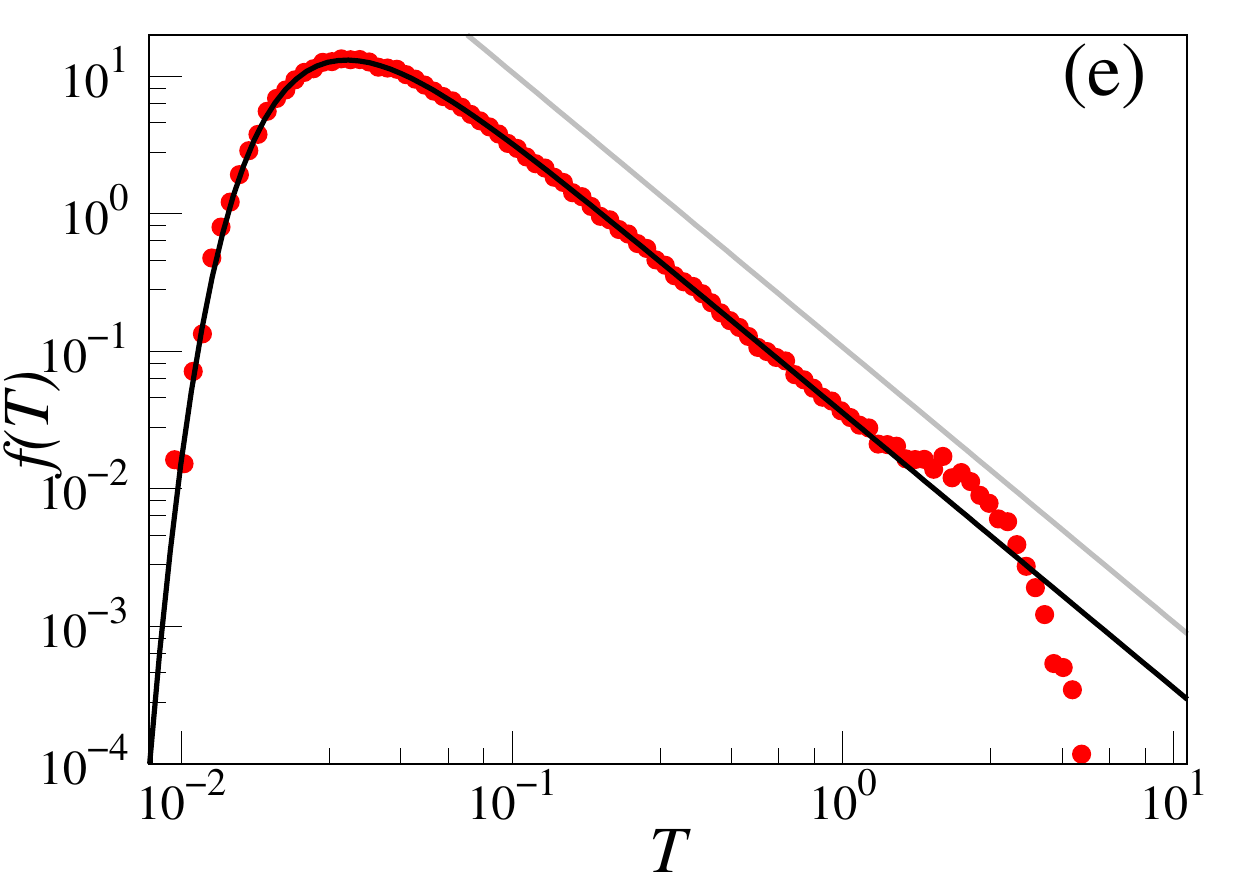}\hspace{0.05\textwidth}\includegraphics[width=0.4\textwidth]{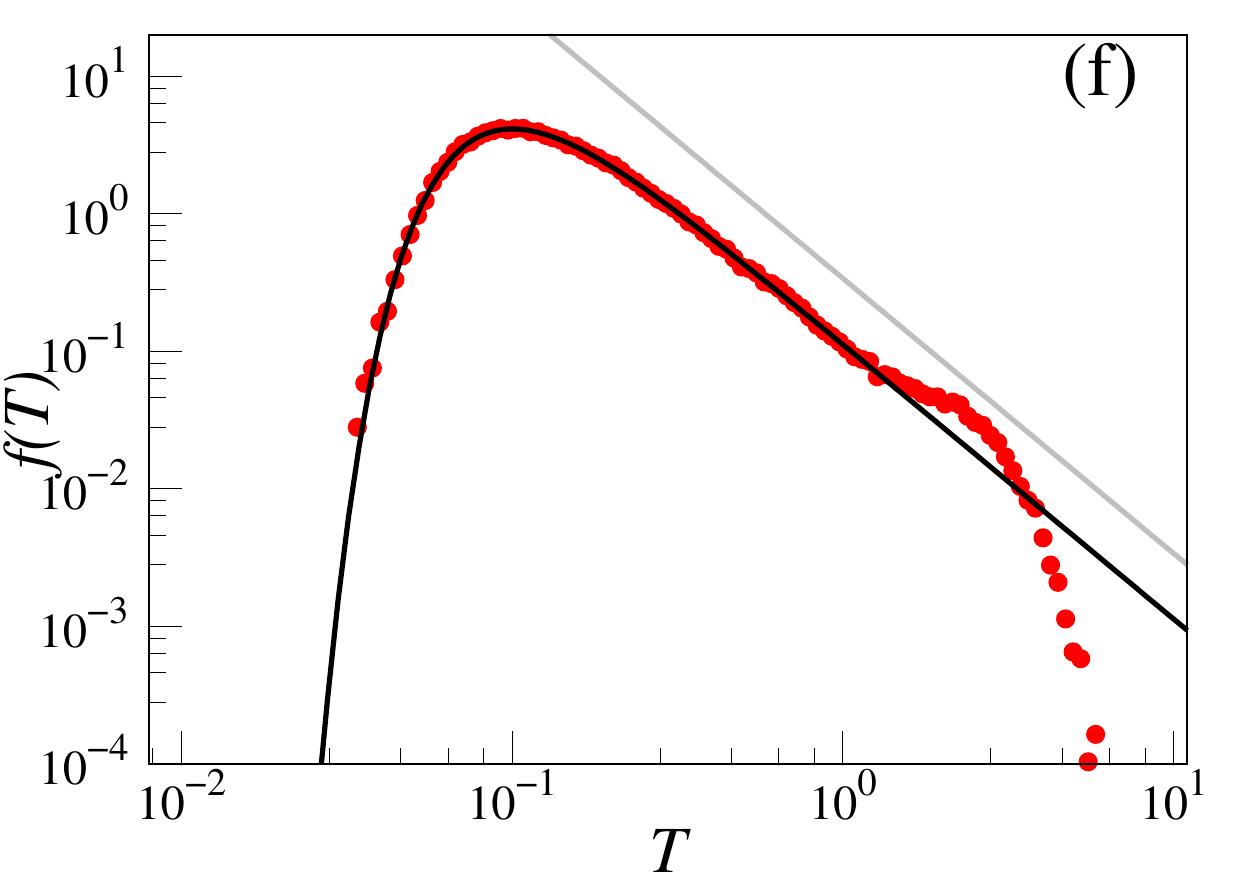}
\caption{The first passage times $T$ distribution (FPTD) of
the scaled voter model, SDE~(\ref{eq:Scaled-Voter}),
for various values of the parameters $x_{0}$ and $\gamma_{s}$ values.
Red points represent the results of numerical simulations.
Black (solid) lines are calculated using an analytical 
Eq.~(\ref{eq:FPTD-x-e1 and 2-thesame and one half}),
grey dashed and grey solid lines show the power-law tail of FPTD $f(t)\sim1/t^{\beta}$
with exponent $\beta=\gamma_{s}/2+1$. The common parameter values were set as follows: $a_{x}=0$  (point of absorption) and $\varepsilon_{1}=\varepsilon_{2}=0.5$.
Initial position is different for the two cases shown:
$x_{0}=10^{-3}$ for (a),(c) and (e), $x_{0}=10^{-2}$ for (b),(d) and (f).
SBM anomalous diffusion exponent is different for the three cases shown:
$\gamma_{s}=1/2$ ($\beta=5/4$) for (a) and (b), $\gamma_{s}=1$ ($\beta=3/2$)
for (c) and (d),
$\gamma_{s}=2$ ($\beta=2$) for (e) and (f).}
 \label{fig3:FPTD-vareps05}
\end{figure}

\begin{figure}
\centering
\includegraphics[width=0.4\textwidth]{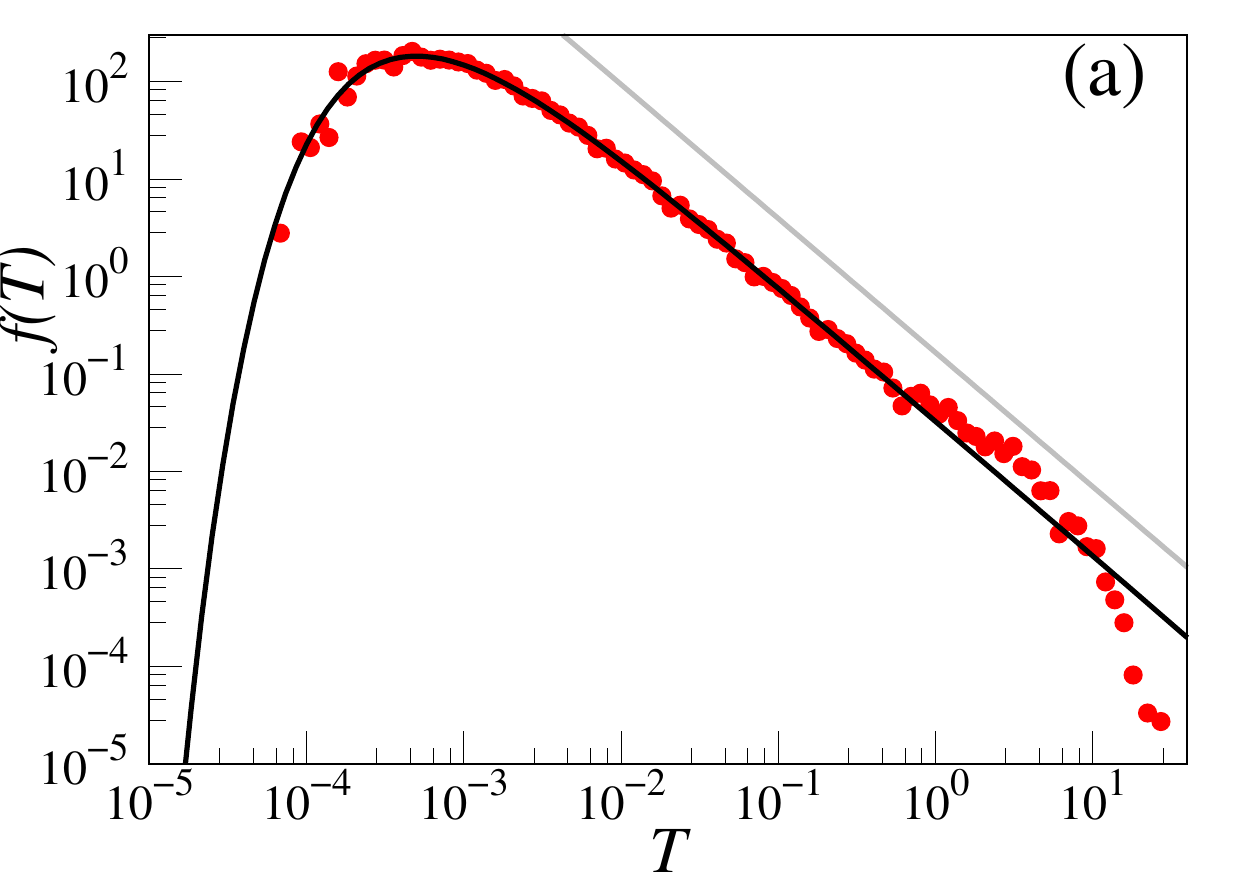}\hspace{0.05\textwidth}\includegraphics[width=0.4\textwidth]{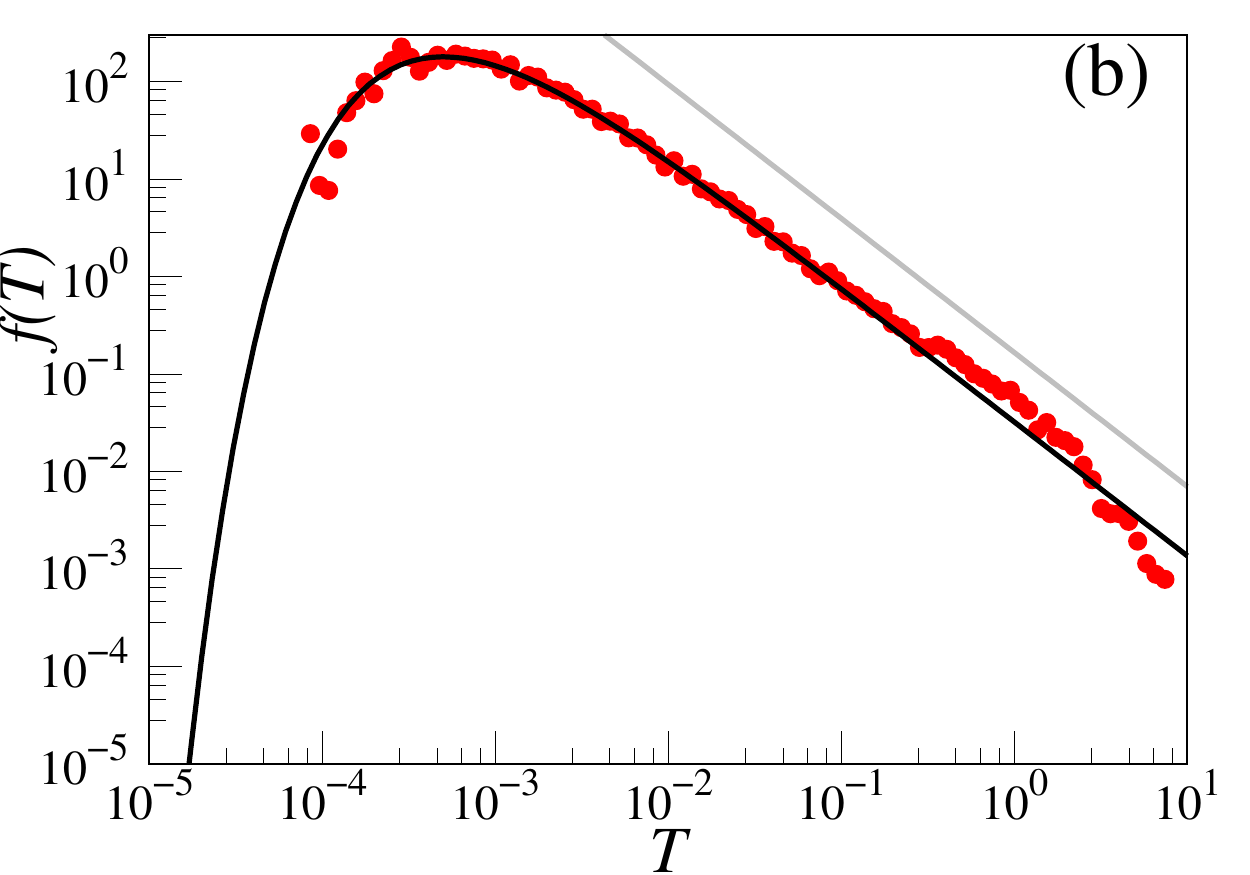}\\\includegraphics[width=0.4\textwidth]{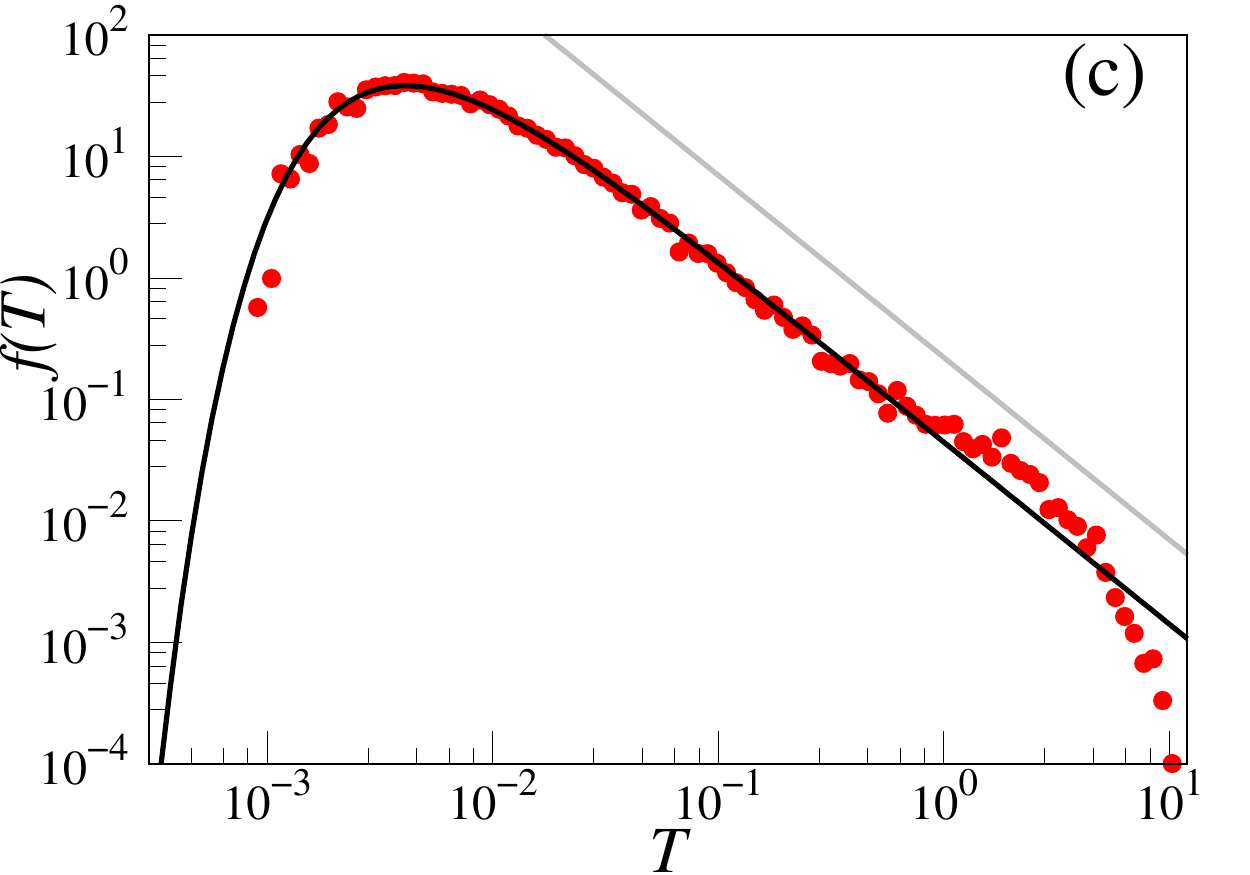}\hspace{0.05\textwidth}\includegraphics[width=0.4\textwidth]{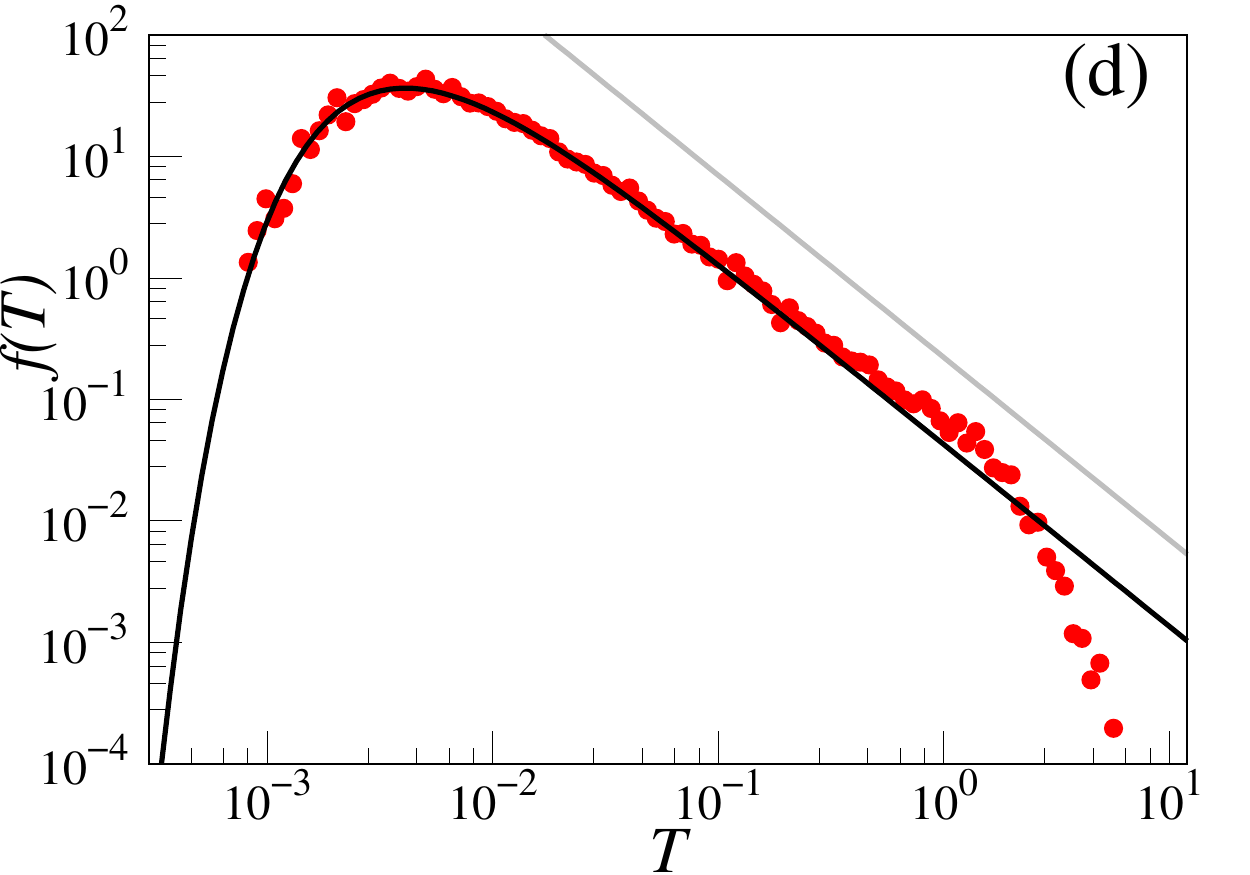}\\\includegraphics[width=0.4\textwidth]{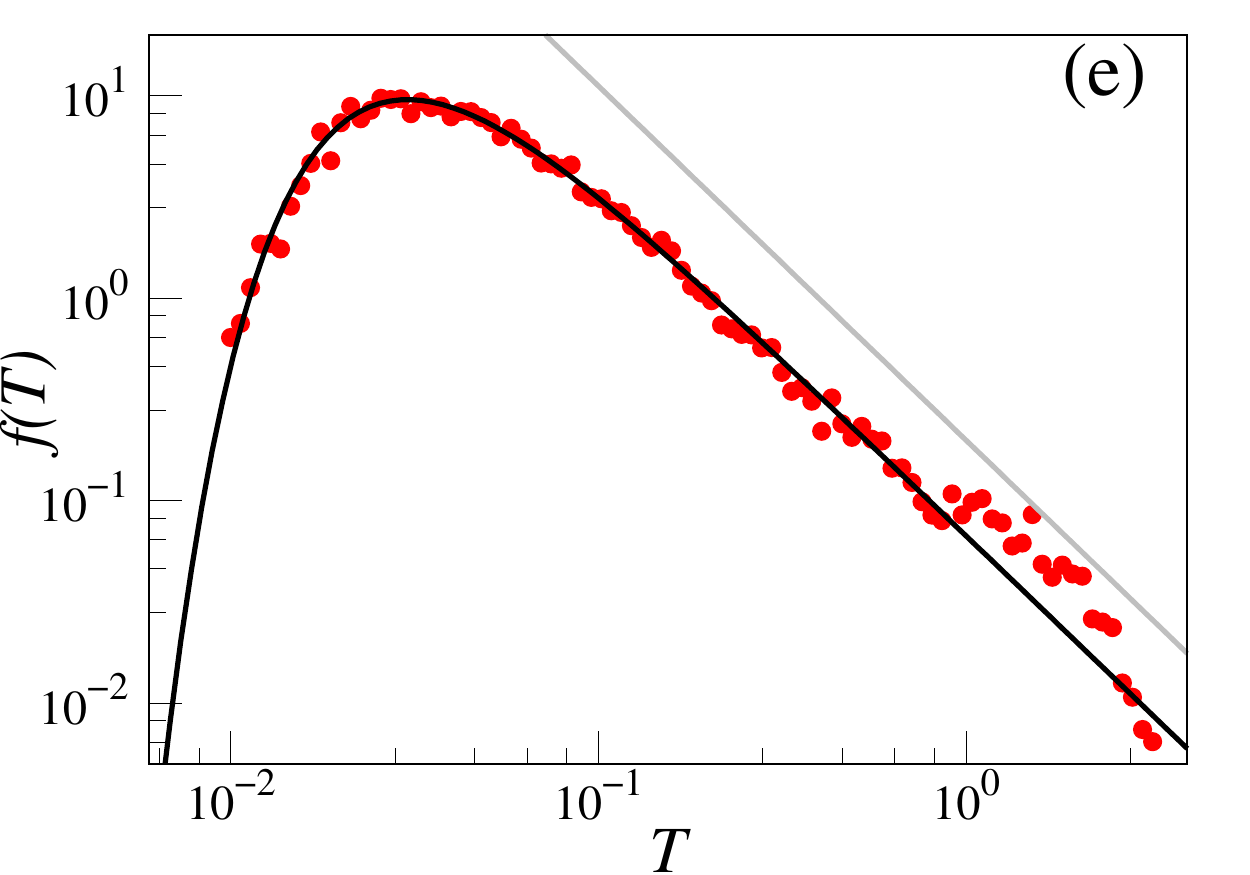}\hspace{0.05\textwidth}\includegraphics[width=0.4\textwidth]{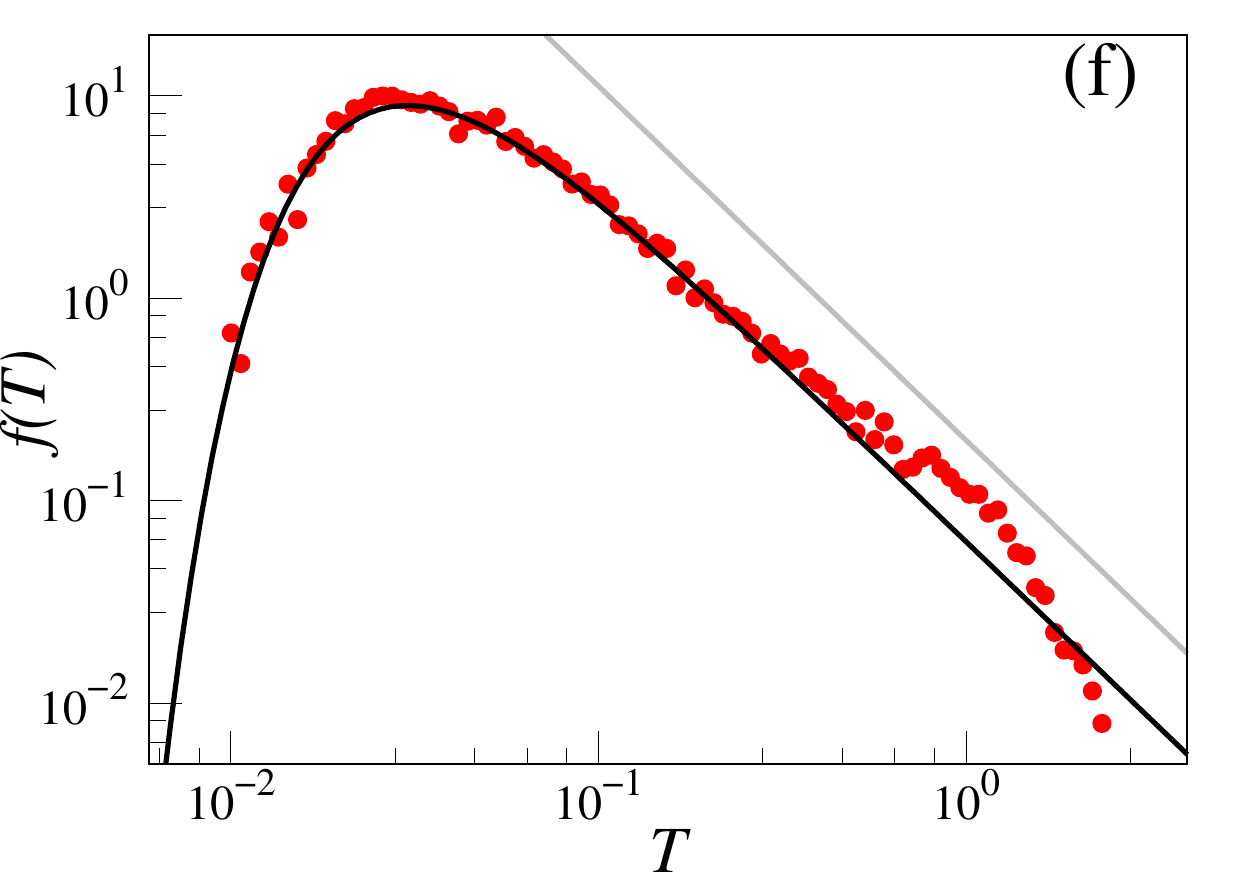}
\caption{The first passage times $T$ distribution (FPTD) of
the scaled voter model, SDE~(\ref{eq:Scaled-Voter}),
for various values of the parameters $\varepsilon$ and $\gamma_{s}$.
Red points represent the results of numerical simulations. 
Black (solid) lines are calculated using an analytical Eq.~(\ref{eq:FPTD-x-e1 and 2-thesame and one half}), and grey solid lines show the power-law tail of FPTD $f(t)\sim1/t^{\beta}$
with exponent $\beta=\gamma_{s}/2+1$. 
The common parameter values were set as follows: $a_{x}=\cos^{2}(1/\sqrt{2})$
($a_{y}=1$), $x_{0}=1/2$ ($y_{0}=\pi/\sqrt{8}$).
The transition rate is different for the two cases shown: $\varepsilon=0.4$
for (a),(c) and (e), $\varepsilon=0.8$ for (b),(d) and (f).
SBM anomalous diffusion exponent is different for the three cases shown:
$\gamma_{s}=3/4$ ($\beta=11/8$) for (a) and (b), $\gamma_{s}=1$
($\beta=3/2$) for (c) and (d), $\gamma_{s}=3/2$ ($\beta=7/4$) for (e) and (f).}
 \label{fig4:FPTD-vareps04-06}
\end{figure}

In this section, we have shown that the noisy voter model
with the time-dependent herding behavior intensity is a nonlinear
transformation of SBM in an external field. 
By using this similarity, we have obtained an analytical approximation of FPTD.
This approximation suggests that in the case of a symmetrical
noisy voter model (with $\varepsilon=1/2$), FPTD has the same power-law
tail as SBM FPTD. To test this prediction, we performed numerical simulations. 
In the numerical simulations, we use a modified next reaction method \cite{Anderson2007JCP,Anderson2011Springer}
with an additional scaling modification, which allows us to improve simulation speed by dynamically
scaling $N$ whenever greater precision is needed (see Appendix~\ref{sec:simulation-method}
for more details).
The numerical simulations confirm that the scaled voter model
exhibits FPTD with a power-law tail whose exponent is well predicted by 
the analytical expressions derived in this section (see Fig.~\ref{fig3:FPTD-vareps05}).
In addition, by using numerical simulations, we have also examined the asymmetric case, FPTD of the scaled voter model still retains the predicted power-law tail exponent. (see Fig.~\ref{fig4:FPTD-vareps04-06}).
The proposed analytical formula Eq.~(\ref{eq:FPTD-x-e1 and 2-thesame and one half})
predicts the overall shape of FPTD quite well up to large times. For
large times, we see a cut-off of the power-law tail. We suspect that
this deviation from the power-law might be due to reflective boundaries
used in the numerical simulations or due to the influence of parameter $\varepsilon$
describing independent transition rates. To explain this phenomenon 
a more precise approximation is needed. 
It would need to take into consideration not only one absorbing
boundary but the combination of the reflective and absorbing boundaries.

FPTD of  the scaled noisy voter model has the
same power-law tail as the SBM FPTD,
but these processes can be differentiated by different power-law scaling
of their MSD. The MSD for SBM is $\langle x^{2}(t)\rangle-\langle x(t)\rangle^{2}\sim t^{\gamma_{s}}$
and for the proposed noisy voter model MSD is $\langle x^{2}(t)\rangle-\langle x(t)\rangle^{2}\sim t^{2\gamma_{s}}$.
When we take into consideration the initial position things can become
a little bit more complicated. For $x_{0}>b/3$ variance has the same
power-law scaling as MSD and processes can be easily separated from each other.
For $x_{0}<b/3$ variance of the proposed model can have double power-law scaling.
For shorter times, proposed model and SBM variances have identical
power-law scaling. For longer times, proposed model variance starts growing with
a double exponent compared to SBM. Therefore, the knowledge
of both FPTD and variance lets us differentiate the considered scaled
voter model from various other long-range memory processes
such as SBM and fBm (fBm and SBM have identical dependence on FPTD
and power-law scaling of MSD). In addition, because the FPTD can have power-law
tails with other exponents than $-3/2$, this lets us differentiate
our model from other nonlinear transformations of the noisy voter
model \cite{Kazakevicius2021PRE} and L\'evy flights.

\section{Scaled voter model relation to other agent-based models}

\label{sec:Applications} In this section, we will study 
how the scaled voter model relates to other
ABMs and stochastic processes. 
We will show the noisy voter model with 
the time-dependent herding behavior intensity arises as 
a special case of other more complicated agent-based models.

In the case of $x_{0}<b/3$, the scaled voter model exhibits double power-law
scaling of variance (see Eq~(\ref{eq:Double power-law})).
For $0.5<\gamma_{s}<1$, we can observe both types of anomalous diffusion:
the subdiffusion transitioning into superdiffusion.
Similar double power-law scaling has been obtained
by performing a numerical simulation of multi-state agent-based models
describing carrier-based transport through a line of cells \cite{Kruse2008}. 
The aforementioned research and other studies  \cite{Metzler2000,Metzler2000JCHempPhys}
motivated us to search  for the relation between the scaled
noisy voter model and multi-state ABMs. In this section,
we present the simplest possible cases.

Let us start with a system of coupled SDEs
\begin{equation}
dn_{1}=n_{2}\left[\varepsilon_{1}\left(1-n_{1}\right)-\varepsilon_{2}n_{1}\right]dt+\sqrt{2n_{2}n_{1}\left(1-n_{1}\right)}dW_{1,t}\, .\label{eq:coupled-1}
\end{equation}
\begin{equation}
dn_{2}=f(n_{1},n_{2})dt+g(n_{1},n_{2})dW_{2,t}\, .\label{eq:coupled-2}
\end{equation}

A special case of the two dimensional stochastic process above 
has been used to model super-exponential
financial bubbles, where the stochastic variable $n_{2}(t)=h(t)$
was interpreted as herding fluctuations
driven by the Ornstein-Uhlenbeck process \cite{Kaizoji2015JEBO}. 
Therefore Eqs.~(\ref{eq:coupled-1}) and (\ref{eq:coupled-2}) systems 
can be interpreted as noisy voter models with modulated
herding behavior intensity $h(t)=n_{2}(t)$ by SDE~\eqref{eq:coupled-2}.
In Ref.~\cite{Ruseckas2016JStatMech} a more general
case of coupled SDEs have been used to model long-range memory process such as Gaussian
$1/f$ noise.

The assumption that herding behavior 
is time-dependent is quite common in the literature \cite{Haas2013NAJEF,Babalos2015AE,Kaizoji2015JEBO,Stavroyiannis2019RBF}.
Yet often it is assumed to be a stochastic process, while here we assume that stochastic fluctuations of variable $n_{2}$ (herding behavior intensity)
can be neglected (condition $g(n_{1},n_{2})=0$ or condition $dW_{2,t}=0$
must be satisfied in SDE~\eqref{eq:coupled-2}). Also, we assume that stochastic variable $n_{2}$
is independent from $n_{1}$. 
These assumptions are needed to introduce long-range 
memory properties into our model.
Our assumption might be correct when the trend
in herding time dependence  is much more significant than the noise.

From the aforementioned assumptions follows $f(n_{1},n_{2})=f(n_{2})$,
and the SDE~\eqref{eq:coupled-2} becomes a deterministic ordinary differential
equation
\begin{equation}
dn_{2}=f(n_{2})dt.\label{eq:n-2-generator}
\end{equation}

Let us consider the influence of $f(n_{2})$ on the equation system. If we set
\begin{equation}
f(n_{2})=(\gamma_{s}-1)\gamma_{s}^{\frac{1}{\gamma_{s}-1}}n_{2}^{1-\frac{1}{\gamma_{s}-1}},
\end{equation}
then herding behavior intensity is growing according to $n_{2}(t)=\big(\gamma_{s}^{\frac{1}{\gamma_{s}-1}}t+n_{20}^{\frac{1}{\gamma_{s}-1}}\big)^{\gamma_{s}-1}$
and SDE~\eqref{eq:coupled-2} can be rearranged into
\begin{equation}
dn_{1}=\left[\varepsilon_{1}\left(1-n_{1}\right)-\varepsilon_{2}n_{1}\right]\big(\gamma_{s}^{\frac{1}{\gamma_{s}-1}}t+n_{20}^{\frac{1}{\gamma_{s}-1}}\big)^{\gamma_{s}-1}dt+\sqrt{2\big(\gamma_{s}^{\frac{1}{\gamma_{s}-1}}t+n_{20}^{\frac{1}{\gamma_{s}-1}}\big)^{\gamma_{s}-1}n_{1}\left(1-n_{1}\right)}dW_{1,t}\,.\label{eq:time-herd-non-linear}
\end{equation}

In the above, $n_{20}$ is the initial herding behavior intensity value at the time
$t=0$. If $t\gg n_{20}^{\frac{1}{\gamma_{s}-1}}/\gamma_{s}^{\frac{1}{\gamma_{s}-1}}$
or we assume that initial herding $n_{20}=0$ (at time moment $t=0$
we have only individualistic behavior), then herding behavior intensity is growing
according to $n_{2}(t)=\gamma_{s}t{}^{\gamma_{s}-1}$ and SDE~\eqref{eq:time-herd-non-linear}
becomes identical to SDE~\eqref{eq:Scaled-Voter} describing 
the scaled voter model.

If $f(n_{2})=a(1-n_{2})$, and we set that initial herding $n_{20}=0$
(at first $t=0$ we have only individualistic behavior $n_{2}=0$)
then herding behavior intensity is growing according $n_{2}=1-e^{-at}$ and
\begin{equation}
dn_{1}=\left[\varepsilon_{1}\left(1-n_{1}\right)-\varepsilon_{2}n_{1}\right](1-e^{-at})dt+\sqrt{2(1-e^{-at})n_{1}\left(1-n_{1}\right)}dW_{1,t}\,.
\end{equation}
For small times $t\ll a^{-1}$, we obtain that stochastic variable
$n_{1}$ satisfies equation
\begin{equation}
dn_{1}=\left[\varepsilon_{1}\left(1-n_{1}\right)-\varepsilon_{2}n_{1}\right]atdt+\sqrt{2atn_{1}\left(1-n_{1}\right)}dW_{1,t}\,.\label{eq:SBM-conection}
\end{equation}

If we change the timescale $t\rightarrow\frac{2t}{\sqrt{a}}$ (or by
set $a=2$), we see that SDE is a special case of the scaled voter model
with $\gamma_{s}=2$. 
SDE~\eqref{eq:SBM-conection} describes a special case of 
the scaled voter model that could be used 
to approximate the short-time dynamics of the Kaizoji model 
\cite{Kaizoji2015JEBO} when the trend in herding time dependence
is much more significant than the noise.

In this section, we made some suggestions on how the scaled noisy voter
model might be applied to analyze other multi-state ABMs. Here, we simply
assumed that only one variable is affected by the noise and the second
variable behaves in a deterministic manner. Therefore the second variable
can be interpreted as a mechanism generating time-dependent herding
behavior intensity. In the general case, to show that two variable voter models
can be approximated by using the scaled voter model one should use adiabatic
or other elimination procedures of one variable \cite{Luszka2005,Hasegawa2007,Kogan2014}.
To perform such a procedure one should know the exact form of the $g(n_{1},n_{2})$
coefficient at the diffusion term. In some cases, 
drift and diffusion coefficients are known \cite{Kaizoji2015JEBO};
in other cases they can be determined from empirical data \cite{Fan2003}.
We are planning to move in this direction for future research.

\section{Conclusions}

\label{sec:concl}

We have shown that a one-dimensional noisy voter model with a time-dependent
herding behavior intensity for short times can be approximated by the CIR process
with time-dependent coefficients. Additionally, general analytical expressions
for the first and second moments, MSD and variance have been obtained. 
In the particular case when the herding behavior intensity is
a power-law function (scaled voter model) of time exact moments and FPTD have been calculated.
The time-dependent herding behavior intensity was chosen in such a form that
the proposed scaled voter model would be a stationary process 
with the same variance scaling and power-law tail in FPTD as fBm. 
Such a tail in FPTD is unique compared 
to other nonlinear transformations of the voter models \cite{Kononovicius2013EPL,Kazakevicius2021PRE}. 
However, the proposed model can not reproduce power-law PSD as nonlinear transformations of voter models \cite{Kononovicius2013EPL,Kazakevicius2021PRE} (or SDEs \cite{Kazakevicius2021Entropy}).

The proposed model has bistable steady-state distribution as bounded fBm \cite{Guggenberger2019,Vojta2020,Kononovicius2022CSF}.
We expect that in the future the combination of the scaled model 
and one of the proposed nonlinear transformations discussed in
Ref.~\cite{Kazakevicius2021PRE} will lead 
to an ABM that mimics all statistical properties of fBm.

One of the long-range memory indicators is power-law scaling of MSD.
In addition, power-law scaling of MSD can be an indication of anomalous
diffusion. From Eq.~(\ref{eq:Double-p-law_Exponent}) it follows
that for times smaller than $t_{c}$ the SDE~(\ref{eq:Voter-small-X})
with power-law $h(t)$ generated signal MSD, $\langle x^{2}(t)\rangle-\langle x(t)\rangle^{2}\sim t^{2\gamma_{s}}$,
scales with doubled exponent compared to SBM. Therefore we can observe anomalous
diffusion for $0<\gamma_{s}<1$, and for $\gamma_{s}>1$, we can observe
superballistic motion.
In addition, we also have time evolution of the variance.
In the case of $x_{0}>b/3$ (here $b=\frac{\varepsilon_{1}}{\varepsilon_{1}+\varepsilon_{2}}$
and $\varepsilon_{i}$ are transition rates) variance exhibits the
same anomalous scaling as SBM up to critical time $t_{c}$ (see Fig.~\ref{fig:mean-var-eps3-x0-03 gg b3}).
After critical time $t_{c}$, moments tend to their steady-state values.
In contrast, other transformations of the noisy voter model lead
to inverse power-law decay of variance from the initial
values to the steady-state value. \cite{Kazakevicius2021PRE}.
In the case
of $x_{0}<b/3$ and $0.5<\gamma_{s}<1$ (see Fig.~\ref{fig:means-veps3-double-p-law}) we
can observe double power-law scaling of variance: 
the subdiffusion transitioning into superdiffusion (see Eq~(\ref{eq:Double power-law})).
Similar double power-law scaling has only been obtained 
in a more complicated two-dimensional models \cite{Metzler2000,Metzler2000JCHempPhys, Barkai2001,Kruse2008}.

In Sec.~\ref{sec:FPTD}, we have shown that the noisy voter model
with the time-dependent herding behavior intensity is a nonlinear
transformation of SBM in an external field. 
By using this similarity analytical approximation for FPTD
was obtained. This approximation suggests that the scaled voter model FPTD 
has the same power-law tail as SBM FPTD. 
The numerical simulations confirm the existence of such a tail 
for a variety of parameter values (see Figs.~\ref{fig3:FPTD-vareps05} and \ref{fig4:FPTD-vareps04-06}).
The derived analytical approximation, Eq.~(\ref{eq:FPTD-x-e1 and 2-thesame
and one half}),
predicts the overall shape of FPTD quite well up to large times. For
large times, we see a cut-off of the power-law tail.
The scaled noisy voter model FPTD has the
same power-law tail as the SBM FPTD,
but these processes can be differentiated by different power-law scaling
of their MSD. The MSD for SBM is $\langle x^{2}(t)\rangle-\langle x(t)\rangle^{2}\sim t^{\gamma_{s}}$
and for the proposed noisy voter model MSD is $\langle x^{2}(t)\rangle-\langle x(t)\rangle^{2}\sim t^{2\gamma_{s}}$. Therefore the knowledge
of both FPTD and variance lets us differentiate the considered scaled
voter model from various other long-range memory processes
such as SBM, fBm (fBm has the same FPTD and MSD as SBM, but different PSD, L\'evy flights or other nonlinear transformations of the noisy voter
model \cite{Kazakevicius2021PRE}.

{\small\singlespacing
\section*{Author contributions}

Conceptualization: R.K.; Methodology: R.K. and A.K.; Software: A.K.;
Writing - Original Draft: R.K. and A.K.; Writing - Review \& Editing: R.K.
and A.K.; Visualization: A.K.}

{\small\singlespacing
\section*{Acknowledgement}

This project was funded by the European Union (Project No.
09.3.3-LMT-K-712-19-0017) under an agreement with the Research Council of
Lithuania (LMTLT).}

\appendix

\section{Numerical simulation of the scaled voter model}

\label{sec:simulation-method}

In this appendix, we briefly discuss the numerical simulation method used
in this paper.

Typically, to numerically simulate the noisy voter model, it is sufficient
to use rejection-based simulation methods or a Gillespie method.
These methods are applicable because the transition rates in the noisy
voter model are not explicitly time-dependent. Here, however, we
have considered the scaled voter model, which has time-dependent
transition rates:
\begin{align}
\pi\left(X\rightarrow X+1,t\right)= & \gamma t^{\gamma-1}\left(N-X\right)\left(\varepsilon_{1}+X\right)=\gamma t^{\gamma-1}\pi_{b},\\
\pi\left(X\rightarrow X-1,t\right)= & \gamma t^{\gamma-1}X\left(\varepsilon_{2}+\left[N-X\right]\right)=\gamma t^{\gamma-1}\pi_{d}.
\end{align}

In the above, $\pi_{b}$ and $\pi_{d}$ gather the terms which are
not time-dependent. Yet some of those terms depend on $X$, which
is not constant and changes as the simulation progresses, though, changes
in $X$ occur during the transitions themselves.

We simulate the model with time-dependent transition rates using
the modified next reaction method \cite{Anderson2007JCP,Anderson2011Springer},
which allows the transition rates to be time-dependent. In general,
the modified next reaction method requires solving multiple integral
equations every time the system state is updated \cite{Anderson2007JCP,Anderson2011Springer}.
This complication is introduced by the time-dependence of the transition
rates, though, in our particular case the required integrals of the
transition rates over time can be calculated analytically:
\begin{equation}
\int_{t}^{t+\tau}\pi\left(X\rightarrow X\pm1,s\right)ds=\gamma\pi_{b,d}\int_{t}^{t+\tau}s^{\gamma-1}ds=\pi_{b,d}\left[\left(t+\tau\right)^{\gamma}-t^{\gamma}\right].
\end{equation}

Obtaining this result allows us to avoid the numerical solution of
integral equations, which speeds up the numerical simulation. See
Algorithm~1 for a detailed description of the employed
numerical simulation algorithm. The algorithm was implemented in C
and the code was made available on GitHub (URL: \url{https://github.com/akononovicius/anomalous-diffusion-in-nonlinear-transformations-of-the-noisy-voter-model}).
\begin{figure}[!h]
\centering
\includegraphics[width=1.0\textwidth]{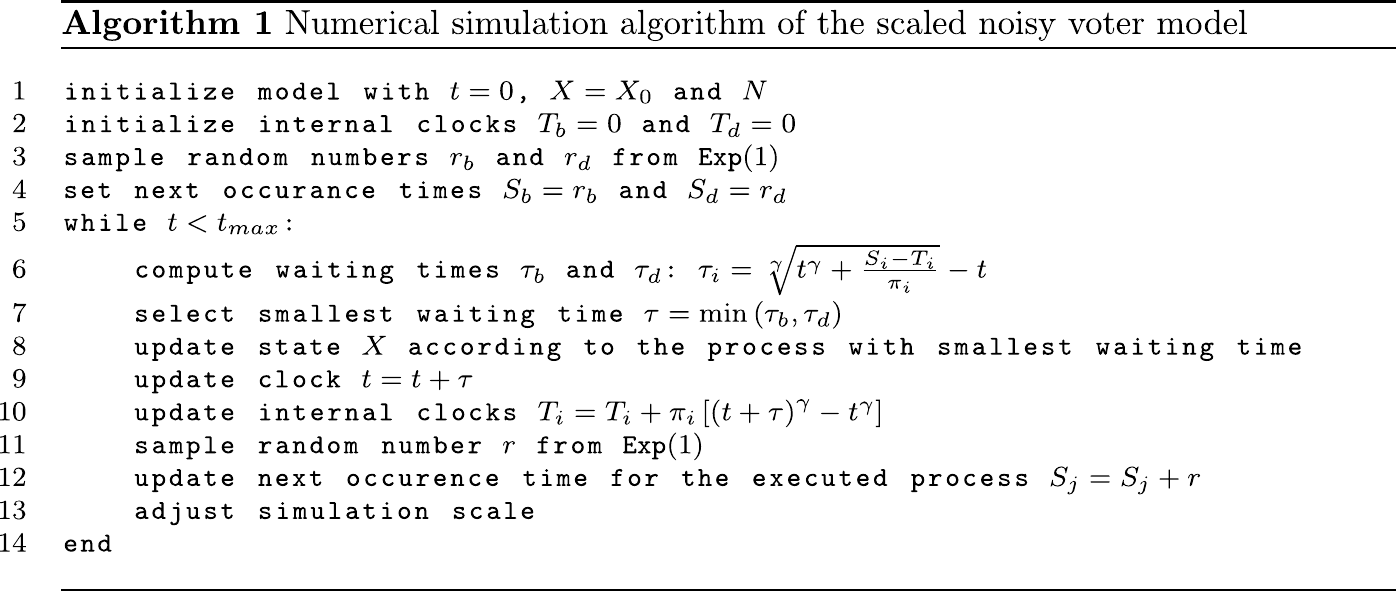}
 \label{alg:mnrm-nvm}
\end{figure}

Note that our implementation also includes dynamic scaling of the
simulation if the system state $X$ gets close to either $0$ or $N$.
Dynamic scaling of the simulation allows us to diminish the observed
discretization effects while keeping the duration of the numerical
simulation reasonable \cite{Kazakevicius2021PRE}. Otherwise, we would
have to increase $N$ for simulation as a whole, which would be quite
costly as the time complexity of the model is $O\left(N^{2}\right)$.
Dynamic scaling allows us to increase $N$ whenever it is necessary,
and otherwise run the model with lower $N$. Unlike in Ref.~\cite{Kazakevicius2021PRE},
here we have implemented both upscaling and downscaling of the simulation
(overall number of particles $N$).

\section{First passage time distribution for scaled Brownian motion with time-dependent drift}

\label{sec:FPTD-for-sigma-mu}

Here we follow the works of Molini \emph{et al.} \cite{Molini2011} and Bhatia
\cite{Bhatia2018}. Let us start with the Fokker-Planck equation with
time-dependent coefficients:
\begin{equation}
\frac{\partial p(x,t|x_{0},0)}{\partial t}=\mu(t)\frac{\partial p(x,t|x_{0},0)}{\partial x}+\frac{1}{2}\sigma^{2}(t)\frac{\partial^{2}p(x,t|x_{0},0)}{\partial x^{2}}.\label{eq:gen-time-D-F-P-eq}
\end{equation}

Here $x=x_{0}$ is the starting point of the process and starting
time is $0$. The PDF satisfies the
initial condition
\begin{equation}
p(x,0|x_{0},0)=\delta(x-x_{0}).\label{eq:Intial-cond}
\end{equation}

We solve this problem with the boundary conditions. At the natural
boundary, the PDF must satisfy
\begin{equation}
p(\infty,t|x_{0},0)=0.
\end{equation}

At absorbing boundary $x=a$ ($a>0$), the PDF must satisfy the Dirichlet boundary
condition:
\begin{equation}
p(a,t|x_{0},0)=0.\label{eq:absorbtion}
\end{equation}

The survival probability $\mathbb{F}(t)$ is defined as the probability
that the process trajectories are not absorbed before time $t$ and the
first passage time density function $f(t)$ is given by the negative
time derivative of the survival probability: 
\begin{equation}
f(t)=-\frac{d}{dt}\mathbb{F}(t).
\end{equation}

The free-space fundamental solution (Green's function) of the Fokker-Planck
equation (\ref{eq:gen-time-D-F-P-eq}) is well-known \cite{Molini2011}:
\begin{equation}
p_{F}(x,t|x_{0},0)=\frac{1}{2\sqrt{\pi
S(t)}}\exp\Big(-\frac{(x-x_{0}-M(t))^{2}}{4S(t)}\Big),
\end{equation}
\begin{equation}
M(t)=\int_{0}^{t}\mu(t^{\prime})dt^{\prime}.\label{eq:M}
\end{equation}
\begin{equation}
S(t)=\frac{1}{2}\int_{0}^{t}\sigma^{2}(t^{\prime})dt^{\prime}.\label{eq:S}
\end{equation}

To solve this problem, we will use the method of
images \cite{Cox1965,Redner2001}: the barrier at $a$ is replaced
by a mirror source located at a generic point $x=m$ (mirror point),
such that the solutions of Eq.~(\ref{eq:gen-time-D-F-P-eq}) emanating
from the original and mirror sources exactly cancel each other at
the absorbing boundary Eq.~(\ref{eq:absorbtion}) at each instant
of time \cite{Redner2001}. This implies the initial conditions in
Eq.~(\ref{eq:Intial-cond}) must now be changed to \cite{Molini2011}
\begin{equation}
p(x,0|x_{0},0)=\delta(x-x_{0})-e^{-\kappa}\delta(x-m),
\end{equation}
where $\kappa$ determines the strength of the mirror image source.
Due to the linearity of the Fokker-Planck equation
(\ref{eq:gen-time-D-F-P-eq}),
a solution for this partial differential equation is provided by 
\begin{equation}
p(x,t|x_{0},0)=p_{F}(x,t|x_{0},0)-e^{-\kappa}p_{F}(x,t|m,0).\label{eq:one-mirror}
\end{equation}

Here we placed an image source $e^{-\kappa}p_{F}(x,t|m,0)$ at $x=m$.
Here $e^{-\kappa}$ is the strength of a mirror. From
Eq.~(\ref{eq:absorbtion}), it
follows that equality
\begin{equation}
p(a,t|x_{0},0)=p_{F}(a,t|x_{0},0)-e^{-\kappa}p_{F}(a,t|m,0)=0,\
\end{equation}
must be true for all times or
\begin{equation}
\exp\Big(-\frac{(a-x_{0}-M(t))^{2}}{4S(t)}\Big)=e^{-\kappa}\exp\Big(-\frac{(a-m-M(t))^{2}}{4S(t)}\Big).
\end{equation}

\begin{equation}
\frac{(a-x_{0}-M(t))^{2}}{4S(t)}=\kappa+\frac{(a-m-M(t))^{2}}{4S(t)}.
\end{equation}

\begin{equation}
(a-x_{0}-M(t))^{2}=4S(t)\kappa+(a-m-M(t))^{2}.\label{eq:Mirror-Point-Equation}
\end{equation}

In general Eq.~(\ref{eq:Mirror-Point-Equation}) is nonsolvable
because we have one equation and two unknown variables $m$ and $\kappa$.
So, we need to make additional assumptions. We require that time-dependent
mean $M(t)$ and variance $S(t)$ at initial
time moment $t_{0}=0$ are also equal to zero (in the case of SBM, this is
true anyway). Therefore at time moment $t=0$, Eq.~(\ref{eq:Mirror-Point-Equation})
becomes 
\begin{equation}
(a-x_{0})^{2}=(a-m)^{2}.
\end{equation}

The aforementioned equation has two solutions:$m=x_{0}$ and $m=2a-x_{0}$. If
we set $m=x_{0}$ from Eq.~(\ref{eq:Mirror-Point-Equation}), it
follows that the mirror is at the same point as the free solution. The mirror
image should mirror the free solution, not copy it. If we set that
$m=2a-x_{0}$, then at the initial time $0$, the free solution and its
mirror image are placed at opposite sites of absorption point $a$
and with the same distance from it as a method of images requires.
By putting the $m$ value into Eq.~(\ref{eq:Mirror-Point-Equation}):
\begin{equation}
(a-x_{0}-M(t))^{2}=4S(t)\kappa+(x_{0}-a-M(t))^{2}.
\end{equation}
\begin{equation}
\frac{\kappa}{x_{0}-a}=\frac{M(t)}{S(t)}.
\end{equation}

For Eq.~(\ref{eq:one-mirror}) to be a solution of the Fokker-Planck
equation (\ref{eq:gen-time-D-F-P-eq}) parameter $\kappa$ must be
a constant, therefore $M(t)$ and $S(t)$ should be chosen such as
\begin{equation}
\frac{M(t)}{S(t)}=q=\mathrm{const},
\end{equation}
then the Fokker-Planck equation (\ref{eq:gen-time-D-F-P-eq}) solution satisfying
an absorbing boundary at $x=a$ is 
\begin{equation}
p(x,t|x_{0},0)=\frac{1}{2\sqrt{\pi S(t)}}e^{-\frac{(x-x_{0}-M(t))^{2}}{4S(t)}}-\frac{1}{2\sqrt{\pi S(t)}}e^{-(x_{0}-a)q}e^{-\frac{(x+x_{0}-2a-M(t))^{2}}{4S(t)}}.\label{eq:One-Mirrow-TPDF}
\end{equation}

The survival probability function $\mathbb{F}(x,t|x_{0},0)$, where
$x=x_{0}$ represents the starting point of the diffusive process,
containing the initial concentration of the distribution and $a$
is a positive lower barrier, such that $a<x_{0}$, is
\begin{equation}
\mathbb{F}(x,t|x_{0},0)=\int_{a}^{\infty}p(x,t|x_{0},0)dx,\quad a<x_{0}.\label{eq:Survival-positive-int}
\end{equation}

Here we show only the derivation of the FPTD if
diffusion is limited to the positive domain of $x$ values $a<x_{0}$
from Eqs.~(\ref{eq:Survival-positive-int}) and (\ref{eq:One-Mirrow-TPDF})
follows:
\begin{equation}
\mathbb{F}(x,t|x_{0},0)=\frac{1}{2}\Bigg(1+\mathrm{Erf}\Big(\frac{x_{0}-a+M(t)}{2\sqrt{S(t)}}\Big)-e^{-(x_{0}-a)q}\bigg[1+\mathrm{Erf}\Big(-\frac{x_{0}-a-M(t)}{2\sqrt{S(t)}}\Big)\bigg]\Bigg).\label{eq:Survival-positive-Gen}
\end{equation}

FPTD $f(t)$ can be obtained by calculating the time derivative of survival
probability 
\[
f(t)=-\frac{\partial}{\partial t}\mathbb{F}(x,t|x_{0},0).
\]
To simplify the derivation, we invoked the previously made assumption that
$q=M(t)/S(t)=\mathrm{const}$:
\begin{equation}
f_{a<x_{0}}(t)=\frac{\Big(x_{0}-a\Big)}{2\sqrt{\pi}}\frac{e^{-\frac{\big(x_{0}-a+M(t)\big)^{2}}{4S(t)}}}{S^{3/2}(t)}\frac{d}{dt}S(t),\label{eq:FPDT-gen-MU-positive}
\end{equation}

To obtain the survival probability function $F(x,t|x_{0},0)$,
when diffusion can occur at negative $x$ domain ( $a>x_{0}$) we
need to calculate integral

\begin{equation}
\mathbb{F}(x,t|x_{0},0)=\int_{-\infty}^{a}p(x,t|x_{0},0)dx\quad a>x_{0}.\label{eq:Survival-negative-int}
\end{equation}

It can be shown that FPTD in such a case is 
\begin{equation}
f_{a>x_{0}}(t)=\frac{\Big(a-x_{0}\Big)}{2\sqrt{\pi}}\frac{e^{-\frac{\big(x_{0}-a+M(t)\big)^{2}}{4S(t)}}}{S^{3/2}(t)}\frac{d}{dt}S(t).
\end{equation}

By comparing $f_{a<x_{0}}(t)$ and $f_{a>x_{0}}(t)$ we see that the
obtained FPTDs differ only in their sign. Therefore, without loss of
generality, we can write
\begin{equation}
f(t)=\frac{|x_{0}-a|}{2\sqrt{\pi}}\frac{e^{-\frac{\big(x_{0}-a+M(t)\Big)^{2}}{4S(t)}}}{S^{3/2}(t)}\frac{d}{dt}S(t).\label{eq:eq:FPDT-gen-MU-SIgma}
\end{equation}

The case when $x_{0}=a$ is trivial if we initially set the process $x$
 at the absorbing boundary: it is absorbed instantly and therefore
FPTD is zero.

Now we will show that the derived general formula can reproduce the results
obtained in the other works \cite{Molini2011}. Therefore we set
$\mu(t)=qAt^{\alpha}$ and $\sigma(t)=\sqrt{2A}t^{\alpha/2}$ as in
Ref. \cite{Molini2011} and put them into Eqs.~(\ref{eq:M}) and (\ref{eq:S})
into (\ref{eq:FPDT-gen-MU-positive}) to obtain
\begin{equation}
f(t)=\frac{|x_{0}-a|(1+\alpha)^{3/2}}{2\sqrt{\pi A}}\frac{1}{t^{\frac{3+\alpha}{2}}}e^{-\frac{\big(qAt^{1+\alpha}+(x_{0}-a)(1+\alpha)\big)^{2}}{4A(1+\alpha)t^{1+\alpha}}}.\label{eq:FPTD-mu-sigma-molini-like}
\end{equation}

If we set $a=0$, FPTD coincides with the well-known result (see (Eq.~(34) in
Ref~\cite{Molini2011}). Parameter
$q$ defines the influence of drift term $\mu(t)$; if we set $q=0$ ($\mu(t)$=0)
we obtain FPTD for the drift-less case 
\begin{equation}
f(t)=\frac{|x_{0}-a|(1+\alpha)^{3/2}}{2\sqrt{\pi A}}\frac{1}{t^{\frac{3+\alpha}{2}}}e^{-\frac{(x_{0}-a)^{2}(1+\alpha)}{4At^{1+\alpha}}}.
\end{equation}

Here in the exponent we have is $(x_{0}-a)^{2}$ in Eq.~(32) in Ref.~\cite{Molini2011} there is a typo: $x_{0}$ that should be $x_{0}^{2}$

\subsection*{Drift-less case}

If we set $\mu=0$ (therefore and $M(t)=0$) then the process is described
by the Fokker-Plank equation 
\begin{equation}
\frac{\partial p(x,t|x_{0},0)}{\partial t}=\frac{1}{2}\sigma^{2}(t)\frac{\partial^{2}p(x,t|x_{0},0)}{\partial x^{2}},\label{eq:time-D-F-P-eq}
\end{equation}
and the method of images leads to a solution satisfying the absorbing boundary
at $x=a$ (see text above)
\begin{equation}
p(x,t|x_{0},0)=\frac{1}{2\sqrt{\pi S(t)}}\bigg(e^{-\frac{(x-x_{0})^{2}}{4S(t)}}-e^{-\frac{(x+x_{0}-2a)^{2}}{4S(t)}}\bigg).\label{eq:eq:One-Mirrow-TPDF-driftless}
\end{equation}

Because there is no drift the mirror strength is 1
($e^{-\kappa}=1,\kappa=0$).
By setting $M(t)=0$ in Eq.~(\ref{eq:eq:FPDT-gen-MU-SIgma}),
we obtain the FPTD for the process 
described by Eq.~(\ref{eq:time-D-F-P-eq}):
\begin{equation}
f(t)=-\frac{\partial}{\partial
t}\mathbb{F}(x,t|x_{0},0)=\frac{|x_{0}-a|}{2\sqrt{\pi}}\frac{e^{-\frac{(x_{0}-a)^{2}}{4S(t)}}}{S^{3/2}(t)}\frac{d}{dt}S(t),\label{eq:FPTD-wiener-Gen-Sigma}
\end{equation}
\[
S(t)=\frac{1}{2}\int_{0}^{t}\sigma(t^{\prime})dt^{\prime}.
\]

\begin{singlespace}

\end{singlespace}

\end{document}